\documentclass[12pt]{iopart}
\usepackage{graphicx} 
\usepackage{xcolor}
\usepackage{hyperref}
\usepackage{caption}
\usepackage{array}
\usepackage{tabularx}
\usepackage{cite}
\usepackage[margin=2.5cm]{geometry}

\begin{document}

\title{Symmetries of Liouvillians of squeeze-driven parametric oscillators}

\author{Francesco Iachello$^{1}$, Colin V Coane$^{1}$ \footnote{Corresponding author} and Jayameenakshi Venkatraman$^{2}$ \footnote{Present address: Department of Physics, University of California, Santa Barbara, Santa Barbara, CA 93106, USA}}

\address{$^{1}$ Center for Theoretical Physics, Sloane Physics Laboratory, Yale University,
New Haven, CT 06520-8120, USA}
\address{$^{2}$ Department of Physics and Applied Physics, Yale University,
New Haven CT 06520, USA}
\eads{\mailto{colin.coane@yale.edu}, \mailto{francesco.iachello@yale.edu}, \mailto{jayavenkat@ucsb.edu}}

\begin{abstract}
We study the symmetries of the Liouville superoperator of one dimensional
parametric oscillators, especially the so-called squeeze-driven Kerr
oscillator, and discover a remarkable quasi-spin symmetry $su(2)$ at integer
values of the ratio $\eta =\omega /K$ of the detuning parameter
$\omega$ to the Kerr coefficient $K$, which reflects the symmetry previously found for
the Hamiltonian operator. We find that the Liouvillian of an $su(2)$
representation $\left\vert j,m_{j}\right\rangle$ has a characteristic
double-ellipsoidal structure, and calculate the relaxation time $T_{X}$ for
this structure. We then study the phase transitions of the Liouvillian which
occur as a function of the parameters $\xi =\varepsilon _{2}/K$ and
$\eta=\omega /K$. Finally, we study the temperature dependence of the spectrum of
eigenvalues of the Liouvillian. Our findings may have applications in the
generation and stabilization of states of interest in quantum computing.
\end{abstract}

\vspace{2pc}

\maketitle

\section{Introduction} \label{intro}

Symmetries of models in which the Hamiltonian operator is expressed in terms of
elements, $\hat{G}_{i}$, of a Lie algebra
$g$, appear in many areas of physics, ranging from
particle physics \cite{gellmann,neeman} to nuclear physics \cite{iac1}
and molecular physics \cite{iac2}. In these models,
 $\hat{H}=\sum_{i}\alpha _{i}\hat{G}_{i}$, where $\alpha _{i}$
are tunable parameters, and the algebra $g$
is called the spectrum generating algebra
(SGA) of the model. Dynamic symmetries occur when
the parameters $\alpha _{i}$ take special values for which $\hat{H}$
becomes a function of the invariant Casimir operators of the algebra
$g$. As a result, its eigenvalues can be
written in explicit analytic form, making a connection between theory and
experiment particularly straightforward. The concept of
dynamic symmetry has been so far applied to closed systems. It is of
importance to see whether or not this concept can be extended to
open quantum systems coupled to an external environment, 
where the role of the Hamiltonian operator is replaced by that of
the Liouvillian superoperator. In this article we address this question and
discover a hitherto unknown symmetry of the Liouvillian of one-dimensional
parametric oscillators.

In recent years, several parametric models have been considered
for possible applications to quantum information science,
such as, among others, the Lipkin-Meshkov-Glick model \cite{lipkin}, the Kerr oscillator model, the
Rabi \cite{rabi} and Dicke \cite{dicke} models and the Jaynes-Cummings
\cite{jaynes} model. In this article, we concentrate our attention to the
squeezed Kerr oscillator, a bosonic model, as a prototype of the class of
models comprising, among others, the Lipkin model \cite{lipkin}, a fermionic
model, and the one-dimensional vibron model \cite{iac2}, a bosonic model.
The study of symmetries of models involving coupled systems of bosons and
fermions, such as the Rabi, Dicke, and Jaynes-Cummings models, will be
deferred to a later publication.

Kerr-nonlinear parametric oscillators (KPOs) can be implemented experimentally
with superconducting quantum circuit oscillators, and their applications to
quantum computation have been considered by many authors
\cite{goto1,goto2,mirrahimi,puri,grimm,blais,darmawan,kwon,frattini,venkatraman1,venkatraman2,kirchmair}.
In a previous publication \cite{iac3}, it was found that the algebraic structure of
the Hamiltonian of the squeezed Kerr oscillator
\begin{equation}
\label{eq:squeezedkerrham}
\hat{H}=-\omega \hat{a}^{\dag }\hat{a}+K\hat{a}^{\dag 2}\hat{a}^{2}-\varepsilon _{2}(\hat{a}^{\dag }\hat{a}^{\dag }+\hat{a}\hat{a})
\end{equation}
where $\hat{a}^{\dag },\hat{a}$ are one-dimensional boson creation and
annihilation operators satisfying $\left[ \hat{a},\hat{a}^{\dag }\right] =1$,
is the symplectic algebra $sp(2,\mathcal{R})\sim su(1,1)$. It was also
found that, for integer values of the ratio $\omega /K\equiv \eta $ in the
Kerr Hamiltonian $\hat{H}=-\omega \hat{a}^{\dag }\hat{a}+K\hat{a}^{\dag 2}\hat{a}^{2}$,
an unexpected dynamic quasi-spin symmetry, $su(2)$, occurs and
that for non-zero values of the ratio $\varepsilon _{2}/K\equiv \xi$ a
Quantum Phase Transition (QPT) \cite{sachdev} and an Excited State Quantum
Phase Transition (ESQPT) \cite{caprio,cejnar1,cejnar2}
occur \cite{prado,chavez}.

In describing Markovian open quantum systems, one needs to go from
a study of the eigenvalues of the Hamiltonian operator to the study of the eigenvalues of the
Liouvillian superoperator, which appears in the Lindblad equation
and governs dissipative dynamics for the system density matrix \cite{lindblad,sudarshan}.
Liouvillians of all parametric oscillators at zero temperature possess $u(1)$
dynamic symmetry, thus eigenvalues can be written in explicit analytic form \cite{iac4,iac5}.
However, it turns out that the Liouvillian of the Kerr oscillator
also has an unexpected $su(2)$ quasi-spin dynamic symmetry
which reflects that found previously for the Hamiltonian operator. 
We thus extend, in the first part of this paper, the concept of dynamic symmetry from
Hamiltonian operators to Liouvillian superoperators. We note that the quasi-spin
dynamic symmetry described here is a ``local'' symmetry in the sense that it
occurs only for certain values of the parameters of the model. This symmetry
differs from the ``global'' symmetry $Z_{2}\equiv \Pi $ (parity) \cite{albert} of the
Liouvillian which occurs for any value in the parameter space.
We also note that these symmetries are different from the so-called ``weak''
symmetries of the Liouvillian \cite{buca} which lead to block-diagonal structures
but do not provide analytic solutions. On the contrary, ``dynamic'' symmetries
provide explicit analytic solutions \cite{iac4,iac5} since, when they occur, the
Hamiltonian for closed systems or the Liouvillian for open systems can be
cast in terms of invariant operators of an algebra, the eigenvalues of which can
be written explicitly in terms of the labels of the irreducible representations of the algebra.

In the second part of the paper, we study QPTs and ESQPTs of open systems,
enlarging the usual definition for closed systems \cite{caprio}. Our
definition for open systems is identical to that proposed in \cite{minganti},
and it considers both the Liouvillian gap \cite{kessler} and the order
parameter. In performing this study we concentrate our attention to the
squeeze-driven Kerr oscillator with quadratic squeezing, as a prototype of
all squeeze-driven bosonic systems, including the squeeze-driven harmonic
oscillator \cite{lieu} and other squeeze-driven fermionic systems which can
be bosonized such as the squeeze-driven Lipkin model \cite{lipkin}. Dissipative
phase transitions are of great importance in a variety of fields, including
photonic quantum systems \cite{houck,fitzpatrick,fink1,rodriguez,fink2,gutierrez},
and the results presented here can be of use for studying these systems.

Finally, in the third part of the paper, we discuss the effects of a non-zero temperature on the
Liouvillian, especially the modification to the eigenvalues of the
truncated harmonic oscillator and the Kerr oscillator. In
particular, we show that the quasi-spin symmetry $su(2)$ persists even at
non-zero temperature, although with modifications. This result is of
particular importance in designing quantum hardware,
since properties of the Liouvillian determine the relaxation rate of the system.

The key result of this paper is the recognition of the quasi
spin-symmetry $su(2)$ of the Liouvillian superoperator, which persists at nonzero squeezing $\xi$
and gives rise to large relaxation times at integer values of the parameter $\eta =\omega /K$.
This result is of use for developing quantum computation devices based on the
Kerr oscillator (KPO) \cite{goto1,goto2}, and generating long-lived states.
Another important development is the
introduction of algebraic methods to the study of open quantum systems
and the derivation in explicit analytic form of solutions
for the eigenvalues of the Liouvillian superoperator.
The algebraic structure of one-dimensional oscillators is relatively simple, but
algebraic methods can play an important role in more
complex situations of coupled oscillators, or oscillators in many dimensions,
as shown in \cite{iac1,iac2} for applications to nuclear and molecular physics.

This paper is structured as follows. We first introduce the
theoretical framework in section~\ref{framework}, derive analytic expressions for the
eigenvalues of parametric one-dimensional oscillators in sections~\ref{spectral1d} and~\ref{zerotempspectrum}, and
confirm analytical formulas with numerical calculations
in section~\ref{numericaleval}.
In section~\ref{kpoquasispin} we introduce and discuss the
$su(2)$ quasispin symmetry of the Kerr oscillator Liouvillian.
In section~\ref{spectraltheorysqueezed}, we consider the squeezed Kerr oscillator, and in
section~\ref{lioueigenvals}, we discuss the
structure of the eigenvalues of the Liouville superoperator. In section~\ref{qpts}, we
study the QPTs that occur as a function of the parameters
$\xi =\varepsilon_{2}/K$ and $\eta =\omega /K$,
first for the Hamiltonian in section~\ref{hamqpts}, and second for the
Liouvillian in section~\ref{liouqpts}, for a fixed value of the ratio $\zeta =\kappa /K$ of
the dissipator $\kappa $ to the Kerr coefficient $K$. We discuss thermodynamic
limits of these QPTs in sections~\ref{kerr2ndorderqpt} and \ref{kerr1storderqpt}. In section~\ref{tempdependence},
we consider temperature dependence, and finally, in section~\ref{conclusions}, we
present our conclusions and indicate directions for future work.

\section{Spectral theory of parametric one-dimensional oscillators} \label{spectraltheory}

\subsection{Theoretical framework} \label{framework}

Consider an open quantum system with Hilbert space $\mathcal{H}$, Hamiltonian $\hat{H}$, and system density matrix $\hat{\rho}(t)$. Assuming the system obeys Markovian dynamics, it can be described by the Lindblad master equation \cite{lindblad,sudarshan,breuer}
\begin{equation}
\label{eq:lindblad1}
\partial _{t}\hat{\rho}(t)=-\rmi\left[ \hat{H},\hat{\rho}(t)\right] +\sum_{i}
\kappa_{i}
\mathcal{D}[\hat{\Gamma}_{i}]\hat{\rho}(t) ,
\end{equation}
where $\hat{H}$ is in units of $\hbar $ and the dissipation superoperator $\mathcal{D}[\hat{\Gamma}_{i}]$ is \cite{lindblad,sudarshan,breuer}
\begin{equation}
\label{eq:dissipator}
\mathcal{D}\left[ \hat{\Gamma}_{i}\right] \hat{\rho}(t)=\hat{\Gamma}_{i}\hat{\rho}(t)\hat{\Gamma}_{i}^{\dag }-\frac{1}{2}
\left[ \hat{\Gamma}_{i}^{\dag }\hat{\Gamma}_{i}\hat{\rho}(t)+\hat{\rho}(t)\hat{\Gamma}_{i}^{\dag }\hat{\Gamma}_{i}\right] 
\end{equation}
Here $\hat{\Gamma}_{i}$ is the Lindblad operator associated
with a specific dissipation channel occurring at a rate
$\kappa _{i} \geq 0$ and describes how the environment acts on the system.
Since the Lindblad equation is linear in $\hat{\rho}$, it can be expressed in terms of the
so-called Liouville superoperator $\mathcal{L}$
\begin{equation}
\label{eq:liouville1}
\partial _{t}\hat{\rho}(t)=\mathcal{L}\hat{\rho}(t)
\end{equation}
which contains an imaginary part describing unitary evolution and a real part characterizing dissipation
\begin{equation}
\label{eq:liouville2}
\mathcal{L}\hat{\rho}(t)=\mathcal{L}^{(0)}\hat{\rho}(t)+\mathcal{L}^{(1)}
\hat{\rho}(t)
\end{equation}
where
\begin{equation}
\label{eq:liouvillesplit}
\mathcal{L}^{(0)}\hat{\rho}(t)=-\rmi\left[ \hat{H},\hat{\rho}(t)\right] ,\qquad \mathcal{L}^{(1)}\hat{\rho}(t)=\sum_{i}
\kappa _{i}
\mathcal{D}\left[ \hat{\Gamma}_{i}\right] \hat{\rho}(t) .
\end{equation}
We have denoted here operators by a hat and superoperators (i.e.~operators of operators) by a script letter.
Superoperators, such as $\mathcal{L}$, acts on the space of linear operators on $\mathcal{H}$,
which we will denote as $\mathcal{H}\otimes \mathcal{H}$. The operator space
$\mathcal{H}\otimes \mathcal{H}$ is itself a Hilbert space, with inner product between
$\hat{A},\hat{B} \in \mathcal{H}\otimes \mathcal{H}$ given by the Hilbert-Schmidt inner product \cite{breuer},
\begin{equation}
\label{eq:hilbertschmidt}
\left\langle \hat{A}, \hat{B} \right\rangle = \mathrm{Tr} \left[\hat{A}^{\dag} \hat{B} \right].
\end{equation}

The eigenvalues of the Liouvillian $\mathcal{L}$ will be denoted by $\lambda_{i} $
\begin{equation}
\label{eq:liouvilleeigs}
\mathcal{L}\hat{\rho}_{i}=\lambda _{i}\hat{\rho}_{i}
\end{equation}
where $\hat{\rho}_{i}$ is the eigenmatrix corresponding to eigenvalue $\lambda _{i}$. Note that here
we have dropped the time dependence for simplicity. The Liouvillian $\mathcal{L}$ is not Hermitian,
so its eigenvalues $\lambda _{i}$ may be complex,
\begin{equation}
\label{eq:lioueigsgeneral}
\lambda _{i}=\rmi\lambda _{i}^{(0)}+\lambda _{i}^{(1)},
\end{equation}
and its eigenmatrices are not necessarily orthogonal,
\begin{equation}
\label{eq:nonorthogonalmatrices}
\left\langle \hat{\rho}_{i}, \hat{\rho}_{j \neq i} \right\rangle = \mathrm{Tr} \left[\hat{\rho}_{i}^{\dag} \hat{\rho}_{j \neq i} \right] \neq 0
\end{equation}

It can be proven \cite{breuer} that $\mathrm{Re}[\lambda _{i}]\leq 0$ $\forall \lambda _{i}$,
and there is always a zero eigenvalue $\lambda _{0}=0$.
Moreover, complex eigenvalues occur in conjugate pairs, as (\ref{eq:liouvilleeigs}) implies
\begin{equation}
\label{eq:conjeigs}
\mathcal{L}\hat{\rho}_{i}^{\dag}=\lambda _{i}^{\ast }\hat{\rho}_{i}^{\dag }.
\end{equation}
\textit{Proof}: Taking the adjoint of (\ref{eq:liouvilleeigs}) and rewriting in terms of (\ref{eq:lindblad1}) gives
\begin{eqnarray}
\label{eq:conjeigproof}
\left(\mathcal{L}\hat{\rho}_{i}\right)^{\dag} &=& \left(-\rmi\left[ \hat{H},\hat{\rho}_{i}\right] \right)^{\dag} + 
\sum_{k} \kappa_{k} \left( \hat{\Gamma}_{k}\hat{\rho}_{i}\hat{\Gamma}_{k}^{\dag } -
\frac{1}{2} \left[ \hat{\Gamma}_{k}^{\dag }\hat{\Gamma}_{k}\hat{\rho}_{i} +
\hat{\rho}_{i}\hat{\Gamma}_{k}^{\dag }\hat{\Gamma}_{k}\right] \right)^{\dag} \nonumber \\
 &=& -\rmi\left[ \hat{H},\hat{\rho}_{i}^{\dag}\right] + 
\sum_{i} \kappa_{i} \hat{\Gamma}_{k}\hat{\rho}_{i}^{\dag}\hat{\Gamma}_{k}^{\dag } -
\frac{1}{2} \left[ \hat{\Gamma}_{k}^{\dag }\hat{\Gamma}_{k}\hat{\rho}_{i}^{\dag} +
\hat{\rho}_{i}^{\dag}\hat{\Gamma}_{k}^{\dag }\hat{\Gamma}_{k}\right] \nonumber \\
 &=& \mathcal{L}\hat{\rho}_{i}^{\dag} = \lambda _{i}^{\ast }\hat{\rho}_{i}^{\dag }.
\end{eqnarray}
Eigenmatrices $\hat{\rho}_{i}$ need not be Hermitian, and if $\hat{\rho}_{i}$ is Hermitian,
the eigenvalue $\lambda_{i} = \lambda _{i}^{\ast}$ must be real. Consequently, real
eigenvalues of degeneracy 1 must have corresponding Hermitian eigenmatrices $\hat{\rho}_{i}$,
and it is always possible to construct Hermitian linear combinations of eigenmatrices for
degenerate real eigenvalues $\lambda_{i}$ of the form $\left( \hat{\rho}_{i}+\hat{\rho}_{i}^{\dag }\right)$
and $\rmi\left( \hat{\rho}_{i}-\hat{\rho}_{i}^{\dag}\right)$.

For computational purposes, it is convenient to work with a vectorized representation of operators and a matrix representation of superoperators \cite{frattini,venkatraman1,albert,minganti,rubio},
where, given a basis $\left\{ \left\vert e_{n}\right\rangle \right\}$ of $\mathcal{H}$,
an operator $\hat{A} \in \mathcal{H}\otimes \mathcal{H}$ is mapped to a vector
\begin{equation}
\label{eq:vectorizedoperator}
\hat{A} = \sum_{n,m} A_{nm} \left\vert e_{n} \right\rangle \left\langle e_{m} \right\vert \rightarrow 
\left\vert A \right\rangle \rangle = \sum_{n,m} A_{nm} \left\vert e_{n} \right\rangle \otimes \left\vert e_{m} \right\rangle .
\end{equation}
From this one can show that left and right multiplication of an operator $\hat{O} \in \mathcal{H}\otimes \mathcal{H}$ on $\hat{A}$ are represented by matrix superoperators as
\begin{equation}
\label{eq:lrmultiplication}
\hat{O} \hat{A} \rightarrow 
\left(\hat{O} \otimes \hat{I} \right) \left\vert A \right\rangle \rangle ,
\qquad
\hat{A} \hat{O} \rightarrow 
\left(\hat{I} \otimes \hat{O}^{T} \right) \left\vert A \right\rangle \rangle ,
\end{equation}
where $\hat{I}$ is the identity. Applying this mapping to the Liouvillian $\mathcal{L}$ yields
\begin{eqnarray}
\label{eq:liouvillevector}
\mathcal{L} &\rightarrow& -\rmi \left[ \left(\hat{H} \otimes \hat{I} \right) - \left(\hat{I} \otimes \hat{H}^{T} \right) \right] \nonumber \\
& & + \sum_{i} \kappa_{i} \left[ \left(\hat{\Gamma}_{i} \otimes \hat{\Gamma}_{i}^{\ast} \right) - 
\frac{1}{2} \left( \hat{\Gamma}_{i}^{\dag }\hat{\Gamma}_{i} \otimes \hat{I} \right) - \frac{1}{2} \left( \hat{I} \otimes \hat{\Gamma}_{i}^{T}\hat{\Gamma}_{i}^{\ast} \right) \right] .
\end{eqnarray}

It is important to note that eigenmatrices of the Liouvillian are not density matrices, since to be a physical density matrix, $\hat{\rho}(t)$ must be Hermitian, positive definite, and have unit trace. However, it is possible \cite{minganti} to construct density matrices from Hermitian linear combinations of eigenmatrices $\hat{\rho}_{i}$.

\subsection{Spectral theory of one-dimensional oscillators} \label{spectral1d}

We consider here one-dimensional oscillators with Hamiltonian written in
terms of one-dimensional creation and annihilation operators $\hat{a}^{\dag},\hat{a}$
with commutation relation $\left[ \hat{a},\hat{a}^{\dag }\right]=1$,
\begin{equation}
\label{eq:oscillatorham}
\hat{H}=\omega _{1}(\hat{a}^{\dag }\hat{a})+\omega _{2}(\hat{a}^{\dag }\hat{a})^{2}+\omega _{3}(\hat{a}^{\dag }\hat{a})^{3}+...=\sum_{k}\omega _{k}(\hat{a}^{\dag }\hat{a})^{k},
\end{equation}
where $\omega _{k}$ are tunable parameters. The eigenvalues of $\hat{H}$ are
trivially given by 
\begin{equation}
\label{eq:oscillatoreigs}
E_{n}=\omega _{1}n+\omega _{2}n^{2}+\omega _{3}n^{3}+..., \qquad n=0,1,2,...
\end{equation}
and the eigenfunctions by 
\begin{equation}
\label{eq:oscillatorkets}
\left\vert n\right\rangle =\frac{1}{\sqrt{n!}}(\hat{a}^{\dag
})^{n}\left\vert 0\right\rangle .
\end{equation}
The Hilbert space $\mathcal{H}$ of this oscillator is infinite-dimensional,
but may be truncated in practical applications to a finite number of bosonic excitations $N$,
such that $n=0,1,...,N$. The model Hilbert space $\mathcal{H}$ of this truncated oscillator 
is of dimension $N_{\rm Fock}=N+1$, and is an invariant subspace of the untruncated, infinite-dimensional space.

The most general spectrum generating algebra for one-dimensional oscillators is the Heisenberg
algebra $h(2)$ \cite{iac5}, composed of operators $\hat{a},\hat{a}^{\dag },\hat{a}^{\dag }\hat{a}$,
and the identity operator $\hat{I}$, with commutation relations 
\begin{eqnarray}
\label{eq:heisenbergalgebra}
\left[ \hat{a},\hat{a}^{\dag }\right] &=&\hat{I}, \quad
\left[ \hat{a},\hat{I}\right] =\left[ \hat{a}^{\dag },\hat{I}\right] =0  \nonumber \\
\left[ \hat{a},\hat{a}^{\dag }\hat{a}\right] &=&\hat{a}, \quad
\left[ \hat{a}^{\dag },\hat{a}^{\dag }\hat{a}\right] =-\hat{a}^{\dag }.
\end{eqnarray}
$h(2)$ is non-compact and its representations are infinite-dimensional. To perform
calculations for truncated oscillators with finite $N$, it is convenient to introduce
an auxiliary boson $s$ \cite{iac5} and introduce operators
$\hat{F}_{-}=\hat{s}^{\dag }\hat{a}$, $\hat{F}_{+}=\hat{a}^{\dag }s$, $\hat{n}=\hat{a}^{\dag }\hat{a}$,
$\hat{n}_{s}=\hat{s}^{\dag }\hat{s}$, such that $\hat{n} + \hat{n}_{s} = \hat{N}$ gives the total boson number $N$.
These four operators satisfy the Lie algebra of $u(2)$, which is therefore a spectrum generating algebra
for the truncated oscillator. By considering only $\hat{n}=\hat{a}^{\dag }\hat{a}$,
one has the algebra of $u(1)$, a subalgebra of $u(2)$ and $h(2)$, often written as
$u(2)\supset u(1)$ or $h(2)\supset u(1)$. Formally, one may obtain the algebra $h(2)$
from $u(2)$ by replacing the operators $\hat{s}$ and $\hat{s}^{\dag }$ by $\sqrt{N}$
and taking the limit $N\rightarrow \infty $. The algebra $h(2)$ is called the contracted algebra of $u(2)$ and denoted by 
\begin{equation}
\label{eq:u2contraction}
u(2)\rightarrow _{c}h(2).
\end{equation}
The Hamiltonian\ $\hat{H}$ in (\ref{eq:oscillatorham}) is written in terms only of
the $u(1)$ Casimir operator, $\hat{n} = \hat{a}^{\dag }\hat{a}$, and thus has a
so-called $u(1)$ ``dynamic symmetry'', that is, a situation in which $\hat{H}$
is written in terms only of invariant operators of an algebra. Note that although for
one-dimensional problems this is a trivial statement, it is not so for
higher dimensional problems as described in \cite{iac5}.

For the open system, we consider bosonic dissipators
\begin{equation}
\label{eq:kerrnonlineardissipator}
\mathcal{D}\left[ \hat{a}^{k}\right] \hat{\rho}(t)=\hat{a}^{k}\hat{\rho}(t)\hat{a}^{\dag k}-\frac{1}{2}
\left[ \hat{a}^{\dag k}\hat{a}^{k}\hat{\rho}(t)+\hat{\rho}(t)\hat{a}^{\dag k}\hat{a}^{k}\right] .
\end{equation}
Particularly important are the linear 
\begin{equation}
\label{eq:kerrlineardissipator}
\mathcal{D}\left[ \hat{a}\right] \hat{\rho}(t)=\hat{a}\hat{\rho}(t)\hat{a}^{\dag }-\frac{1}{2}
\left[ \hat{a}^{\dag }\hat{a}\hat{\rho}(t)+\hat{\rho}(t)\hat{a}^{\dag }\hat{a}\right] 
\end{equation}
and quadratic 
\begin{equation}
\label{eq:kerrquadraticdissipator}
\mathcal{D}\left[ \hat{a}^{2}\right] \hat{\rho}(t) =\hat{a}^{2}\hat{\rho}(t)\hat{a}^{\dag 2}-\frac{1}{2}
\left[ \hat{a}^{\dag 2}\hat{a}^{2}\hat{\rho}(t)+\hat{\rho}(t)\hat{a}^{\dag 2}\hat{a}^{2}\right] 
\end{equation}
dissipators, which may describe one- and two-photon losses to the environment
\cite{frattini,venkatraman1,venkatraman2}.
We note here that the dissipators are also built in terms of elements of $h(2)$
and their powers. The powers of algebra elements also form an algebra,
called the enveloping algebra of $h(2)$. However, one only needs the representation
theory of $h(2)$ to make use of its enveloping algebra in calculations.

In order to find the eigenvalues of $\mathcal{L}$ for these models, one must construct
a basis for its eigenmatrices $\hat{\rho}_{i}$.
A generic operator for a one-dimensional bosonic system
can be expanded onto a Fock basis $\left\{ \left\vert n\right\rangle \right\}$ of oscillator eigenfunctions,
\begin{equation}
\label{eq:kerrdensity}
\hat{\rho}=\sum_{n,m} \rho_{nm} \left\vert n\right\rangle \left\langle m\right\vert
\end{equation}
where $\rho_{nm}$ are its matrix elements,
and we have dropped the time dependence of $\rho_{nm}$ for simplicity.
$\hat{\rho}$ is an element of the operator Hilbert space $\mathcal{H}\otimes \mathcal{H}$,
which for the truncated oscillator, has dimension $(N+1)\times (N+1)=N_{\rm Fock}\times N_{\rm Fock}$.

When there is a $u(1)$ dynamic symmetry, the Hamiltonian is diagonal in the basis
$\left\{ \left\vert n\right\rangle \right\}$, with
eigenvalues $E_{n}$. Hence, the imaginary part of the Liouvillian $\mathcal{L
}^{(0)}\hat{\rho}=-\rmi\left[ \hat{H},\hat{\rho}\right] $ is also diagonal, as one
can see by expanding $\left[ \hat{H},\hat{\rho}\right] =\hat{H}\hat{\rho}-
\hat{\rho}\hat{H}$ and using $\hat{\rho}$ from (\ref{eq:kerrdensity}).
Thus, $\mathcal{L}^{(0)}$ has eigenmatrices of the form
$\left\vert n\right\rangle \left\langle m\right\vert$ and spectrum given by
\begin{equation}
\label{eq:kerrimagliouvillian}
\mathcal{L}^{(0)} \left\vert n\right\rangle \left\langle m\right\vert =-\rmi(E_{n}-E_{m}) \left\vert n\right\rangle \left\langle m\right\vert .
\end{equation}
We remark that $\mathcal{L}^{(0)}$ simply describes the closed system
with no dissipation. Introducing linear dissipation with strength $\kappa$, the Liouvillian
at zero temperature is $\mathcal{L} = \mathcal{L}^{(0)} + \mathcal{L}^{(1)}$, with dissipator
\begin{equation}
\label{eq:zerotempdissipator}
\mathcal{L}^{(1)}\hat{\rho}=\kappa \left( \hat{a}\hat{\rho}\hat{a}^{\dag }-
\frac{1}{2}(\hat{a}^{\dag }\hat{a}\hat{\rho}+\hat{\rho}\hat{a}^{\dag }\hat{a})\right) .
\end{equation}
In the Fock basis, we find the Liouvillian, $\mathcal{L} = \mathcal{L}^{(0)} + \mathcal{L}^{(1)}$,
has matrix elements given by
\begin{eqnarray}
\label{eq:lioumatels}
\mathcal{L} \left\vert n\right\rangle \left\langle m\right\vert &=& 
\left[ -\rmi\left( E_{n}-E_{m}\right) -\frac{\kappa }{2}(n+m) \right]
\left\vert n\right\rangle \left\langle m\right\vert \nonumber \\
& & + \left[ \kappa \sqrt{n}\sqrt{m} \right] \left\vert n-1\right\rangle \left\langle m-1\right\vert
\end{eqnarray}
Making use of the vectorized representation (\ref{eq:vectorizedoperator}),
we note that the matrix representative of $\mathcal{L}$ is upper triangular in the Fock basis, 
\begin{equation}
\label{eq:triangularmatrix}
\left(
\matrix{ \lambda_{00} & \times & & \cr
 & \lambda_{01} & \times & \cr
 &  & \lambda_{02} & \times \cr
0 &  &  & \ddots}
\right)
\end{equation}
with eigenvalues given by its diagonal elements, which are therefore
\begin{equation}
\label{eq:zerotempeigs}
\lambda _{n,m}=-\rmi\left( E_{n}-E_{m}\right) -\frac{\kappa }{2}(n+m).
\end{equation}
Note that the real part of $\lambda _{n,m}$ is always $<0$ and there is a zero
eigenvalue $n=m=0$. Also, eigenmatrices $\hat{\rho}_{n,m} \neq \left\vert n\right\rangle \left\langle m\right\vert$ due to off-diagonal matrix elements of $\mathcal{L}$, however, we are not concerned with explicit expressions of $\hat{\rho}_{n,m}$ here.

As another example, one can treat the Liouvillian at zero-temperature with quadratic dissipation
similarly, $\mathcal{L} = \mathcal{L}^{(0)} + \mathcal{L}^{(1)}$, where
\begin{equation}
\label{eq:zerotempquaddiss}
\mathcal{L}^{(1)}\hat{\rho}=\kappa_{2} \left( \hat{a}^{2} \hat{\rho}\hat{a}^{\dag 2}-\frac{1}{2}(\hat{a}^{\dag 2}\hat{a}^{2} \hat{\rho}+\hat{\rho}\hat{a}^{\dag 2}\hat{a}^{2})\right)
\end{equation}
and $\kappa_{2}$ is the strength of dissipation. Following the same treatment as above for linear dissipation, the Liouvillian has Fock basis matrix elements
\begin{eqnarray}
\label{eq:lioumatelsquad}
\mathcal{L} \left\vert n\right\rangle \left\langle m\right\vert &=& 
\left[ -\rmi\left( E_{n}-E_{m}\right) -\frac{\kappa_{2} }{2}\left( n(n-1) + m(m-1) \right) \right]
\left\vert n\right\rangle \left\langle m\right\vert \nonumber \\
& & + \left[ \kappa_{2} \sqrt{n(n-1)}\sqrt{m(m-1)} \right] \left\vert n-2\right\rangle \left\langle m-2\right\vert
\end{eqnarray}
and eigenvalues
\begin{equation}
\label{eq:zerotempeigsquad}
\lambda _{n,m}=-\rmi\left( E_{n}-E_{m}\right) -\frac{\kappa_{2} }{2}\left[ n(n-1) + m(m-1) \right]
\end{equation}
In principle one may follow this procedure for linear combinations of dissipators of arbitrary $k$-photon losses, (\ref{eq:kerrnonlineardissipator}), but in this article we will focus on the linear case.

In section~\ref{tempdependence} we will also consider non-zero temperatures parametrized in terms
of an average thermal population $\bar{n}_{\rm th}$. The linear dissipator at non-zero temperature
is \cite{frattini,venkatraman1,venkatraman2}
\begin{equation}
\label{eq:nonzerotempdissipator}
\mathcal{L}^{(1)}\hat{\rho}=\kappa \left( 1+\bar{n}_{\rm th}\right) \mathcal{D} \left[ \hat{a} \right] \hat{\rho}+\kappa \bar{n}_{\rm th}\mathcal{D} \left[ \hat{a}^{\dag } \right] \hat{\rho}.
\end{equation}
The new term here is 
\begin{eqnarray}
\label{eq:creationdissipator}
\mathcal{D}\left[ \hat{a}^{\dag } \right] &=&\hat{a}^{\dag }\hat{\rho}\hat{a}-\frac{1}{2}\left( \hat{a}\hat{a}^{\dag }\hat{\rho}+\hat{\rho}\hat{a}\hat{a}^{\dag}\right)  \nonumber \\
&=&\hat{a}^{\dag }\hat{\rho}\hat{a}-\frac{1}{2}\left[ \left( 1+\hat{a}^{\dag
}\hat{a}\right) \hat{\rho}+\hat{\rho}\left( 1+\hat{a}^{\dag }\hat{a}\right) \right] .
\end{eqnarray}
The combined linear dissipator at non-zero temperature, can then be written as
\begin{eqnarray}
\label{eq:nonlineardissipatorfull}
\mathcal{L}^{(1)}\hat{\rho} &=&\kappa \left( \hat{a}\hat{\rho}\hat{a}^{\dag }-\frac{1}{2}\left( \hat{a}^{\dag }\hat{a}\hat{\rho}+\hat{\rho}\hat{a}^{\dag }\hat{a}\right) \right)  \nonumber \\
&&+\kappa \bar{n}_{\rm th}\left( \hat{a}\hat{\rho}\hat{a}^{\dag }+\hat{a}^{\dag }\hat{\rho}\hat{a}\right) -\kappa \bar{n}_{\rm th}\left( \hat{a}^{\dag }\hat{a}\hat{\rho}+\hat{\rho}\hat{a}^{\dag }\hat{a}+\hat{\rho}\right) .
\end{eqnarray}
The full Liouvillian, $\mathcal{L} = \mathcal{L}^{(0)} + \mathcal{L}^{(1)}$, has matrix elements
\begin{eqnarray}
\label{eq:lioumatnonzero}
\mathcal{L} \left\vert n\right\rangle \left\langle m\right\vert &=& 
\left[ -\rmi\left( E_{n}-E_{m}\right) -\frac{\kappa }{2}(1+2\bar{n}_{\rm th})(n+m)-\kappa \bar{n}_{\rm th} \right]
\left\vert n\right\rangle \left\langle m\right\vert \nonumber \\
& & + \left[ \kappa (1+ \bar{n}_{\rm th}) \sqrt{n}\sqrt{m} \right] \left\vert n-1\right\rangle \left\langle m-1\right\vert \nonumber \\
& & + \left[ \kappa \bar{n}_{\rm th} \sqrt{n+1}\sqrt{m+1} \right] \left\vert n+1\right\rangle \left\langle m+1\right\vert
\end{eqnarray}
The matrix representative of $\mathcal{L}$ is now tridiagonal,
\begin{equation}
\label{eq:tridiagonalmatrix}
\left(
\matrix{ L_{00} & \times & & \cr
\times & L_{11} & \times & \cr
 & \times  & L_{12} & \times \cr
 &  & \times  & \ddots}
\right)
\end{equation}
and has
eigenvalues which generally must be computed numerically.

\subsection{Specific cases at zero temperature} \label{zerotempspectrum}

In the following sections, our analysis will focus on the zero temperature case,
$\bar{n}_{\rm th} = 0$, for simplicity and clarity. We will consider non-zero
temperatures, $\bar{n}_{\rm th} \neq 0$, in section~\ref{tempdependence}.

\subsubsection{Harmonic oscillator.} \label{zerotempqho}

The algebraic Hamiltonian is 
\begin{equation}
\label{eq:harmonicoscham}
\hat{H}=\omega \left( \hat{a}^{\dag }\hat{a}\right) =\omega \hat{n}
\end{equation}
with eigenvalues
\begin{equation}
\label{eq:harmonicosceigs}
E_{n}=\omega n 
\end{equation}
The eigenvalues of the Liouvillian superoperator are
\begin{equation}
\label{eq:harmonicoscliouvillian}
\lambda _{n,m}=-\rmi\omega \left( n-m\right) -\frac{\kappa }{2}(n+m).
\end{equation}

\subsubsection{Kerr oscillator.} \label{zerotempkpo}

The Kerr oscillator has gained recent attention for its potential applications to
quantum computing \cite{goto2,frattini,venkatraman1}. Its algebraic Hamiltonian is 
\begin{equation}
\label{eq:kerroscham}
\hat{H}=-\omega \hat{a}^{\dag }\hat{a}+K\hat{a}^{\dag 2}\hat{a}^{2}=-\omega 
\hat{n}+K\hat{n}(\hat{n}-1)
\end{equation}
with eigenvalues
\begin{equation}
\label{eq:kerrosceigs}
E_{n}=-\omega n+Kn(n-1).
\end{equation}
The parameter $\omega $ is denoted by $\Delta $ in \cite{frattini,venkatraman1}.
It is convenient to introduce a dimensionless Hamiltonian
\begin{equation}
\label{eq:dimensionlesskerrham}
\frac{\hat{H}}{K}=-\frac{\omega }{K}\hat{n}+\hat{n}(\hat{n}-1)=-\eta \hat{n}+\hat{n}(\hat{n}-1)=-\eta ^{\prime }\hat{n}+\hat{n}^{2}.
\end{equation}
with eigenvalues
\begin{equation}
\label{eq:dimlesskerreigs}
E_{n}=-\eta^{\prime }n+n^{2}.
\end{equation}
The eigenvalues of the Liouvillian of the Kerr oscillator are
\begin{equation}
\label{eq:kerrlioueigs}
\lambda _{n,m}=-\rmi\left[ \left( -\eta ^{\prime }n+n^{2}\right) -\left( -\eta
^{\prime }m+m^{2}\right) \right] -\frac{\kappa }{2}(n+m).
\end{equation}
The Hamiltonian of the Kerr oscillator has a quasi-spin symmetry for integer
values of the parameter $\eta ^{\prime }=\eta +1$ \cite{iac3} which will be
reflected into symmetries of the eigenvalues of the Liouvillian
superoperator to be discussed in the following subsections.

\subsection{Numerical evaluation of the eigenvalues} \label{numericaleval}

In order to evaluate numerically the eigenvalues of the Liouville superoperator,
one must use the vectorized representation (\ref{eq:vectorizedoperator}-\ref{eq:liouvillevector})
and diagonalize $\mathcal{L}$ in a truncated $N_{\rm Fock}\times N_{\rm Fock}$ dimensional space.
Diagonalization gives complex eigenvalues
\begin{equation}
\label{eq:eigsrealimag}
\lambda _{i}=\mathrm{Re}[\lambda _{i}]+\rmi \cdot \mathrm{Im}[\lambda_{i}].
\end{equation}
which can be displayed as a scatterplot in the complex plane.
We consider here the harmonic oscillator
(\ref{eq:harmonicoscham}) and the Kerr oscillator (\ref{eq:kerroscham}) with linear dissipation.
In all subsequent figures we use a strength of dissipation $\kappa =0.1$ and a value of
$N_{\rm Fock}$ as indicated in each figure.

\subsubsection{Harmonic oscillator.} \label{numericalqho}

\begin{figure}[ht!]
    \centering
    \includegraphics[width=0.5\textwidth]{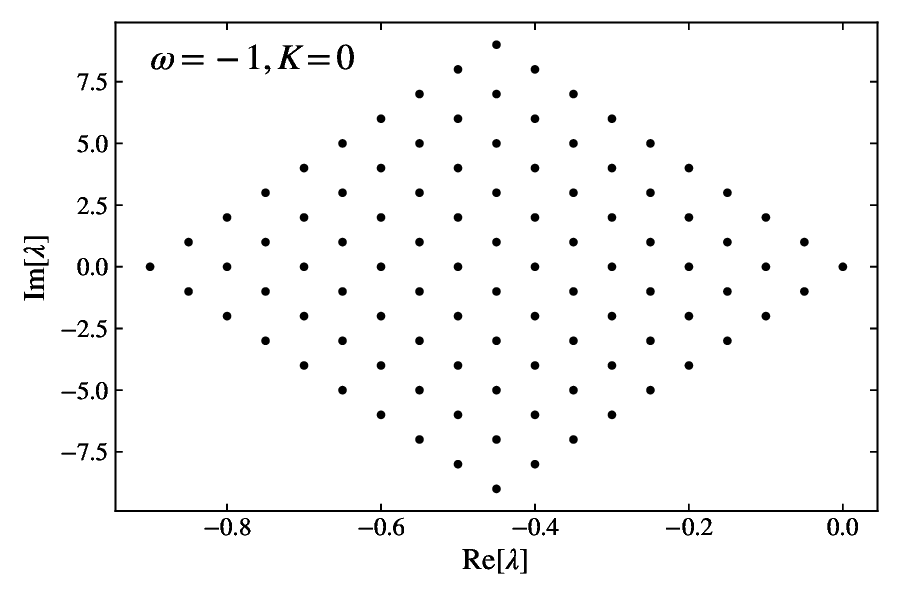}
    \caption{Scatterplot of the harmonic oscillator ($\omega =-1,K=0$) Liouvillian eigenvalues in the complex plane for $N_{\rm Fock}=10$.}
    \label{fig:fig1}
\end{figure}

The Liouvillian spectrum for $\omega =-1,K=0$ is shown in figure \ref{fig:fig1}.

The scatterplot of the harmonic oscillator eigenvalues has two symmetries, reflection
symmetry about the real axis $n\leftrightarrow m$, and reflection about the
imaginary axis at $-\kappa \frac{N}{2}=-\kappa \frac{N_{\rm Fock}-1}{2}$. While
the former occurs for any size of the Hilbert space $\mathcal{H}$
due to eigenvalues occurring in conjugate pairs (\ref{eq:conjeigs}),
the latter depends on the size $N_{\rm Fock}=N+1$ of the Hilbert space.

As discussed in subsection~\ref{spectral1d}, the algebraic structure of the harmonic
oscillator for finite $N$ is $u(2)\supset u(1)$. Bosonic representations
of $u(2)$ are totally symmetric representations characterized by the integer $N$, while those of $u(1)$ are
characterized by the integer $n$, with $n=0,1,2,...,N$, written symbolically
as \cite{iac5}
\begin{equation}
\label{eq:algebraickets}
\left\vert 
\begin{array}{ccc}
\vspace{-0.2em}
u(2) & \supset & u(1) \\ 
\vspace{-0.2em}
\downarrow &  & \downarrow \\ 
\lbrack N] &  & n
\end{array}
\right\rangle ,\qquad n=0,1,2,...,N .
\end{equation}
Numerical diagonalization confirms the analytic formula (\ref{eq:harmonicoscliouvillian}) obtained by
using the dynamic symmetry of the harmonic oscillator.
From (\ref{eq:harmonicoscliouvillian}) one can also see that the geometric
reflection symmetry of the scatterplot about the imaginary axis changes $n,m$
into $n\rightarrow N-m,m\rightarrow N-n$.

\subsubsection{Kerr oscillator.} \label{numericalkpo}

\begin{figure}[ht!]
    \centering
    \includegraphics[width=0.7\textwidth]{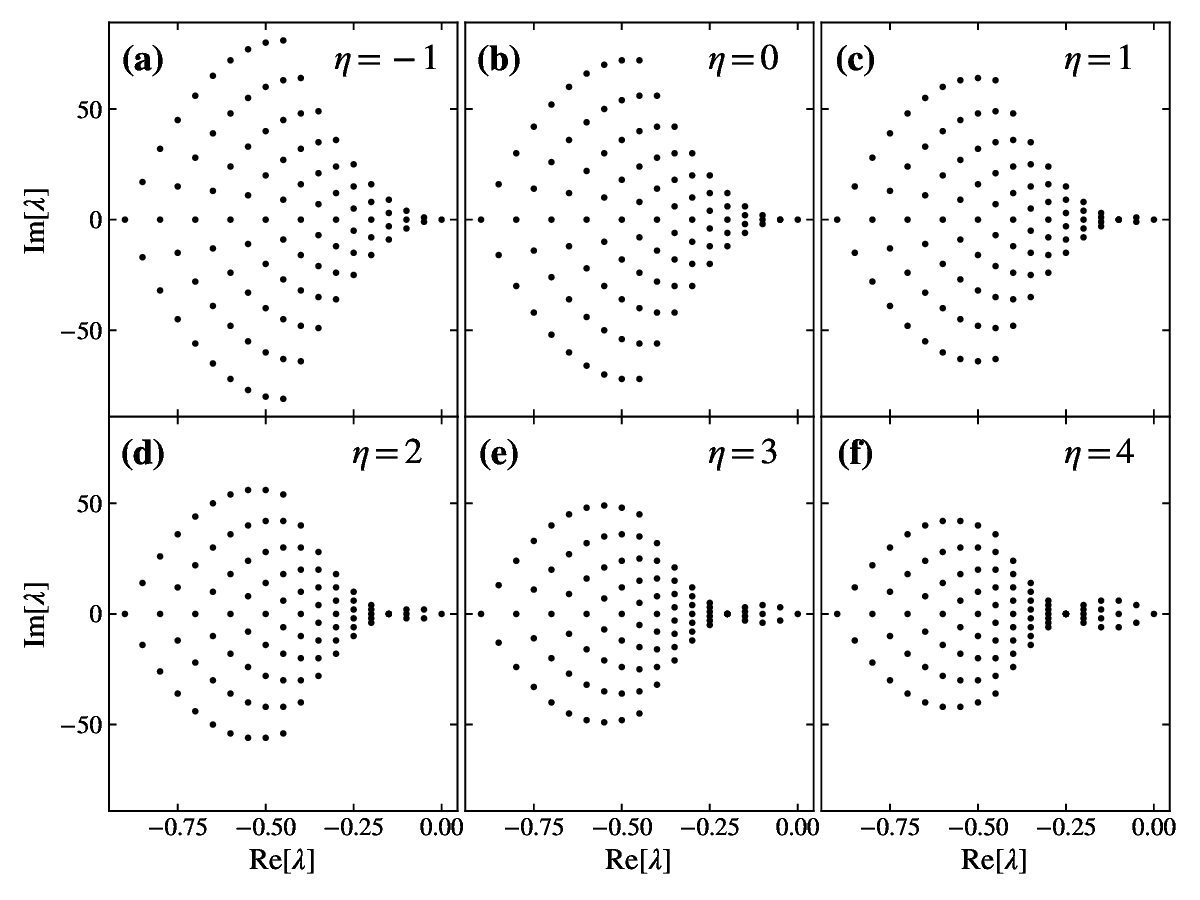}
    \caption{Scatterplots of the Kerr oscillator Liouvillian spectrum for integer values of the ratio
$\omega /K = \eta =-1,0,1,2,3,4$ for $N_{\rm Fock}=10$.}
    \label{fig:fig2}
\end{figure}

Liouvillian spectrum scatterplots for $\omega =-1,0,1,2,3,4$ , $K=1$ for $N_{\rm Fock}=10$
are shown in figure \ref{fig:fig2}, and agree with the analytic formula (\ref{eq:kerrlioueigs}).

\begin{figure}[ht!]
    \centering
    \includegraphics[width=0.7\textwidth]{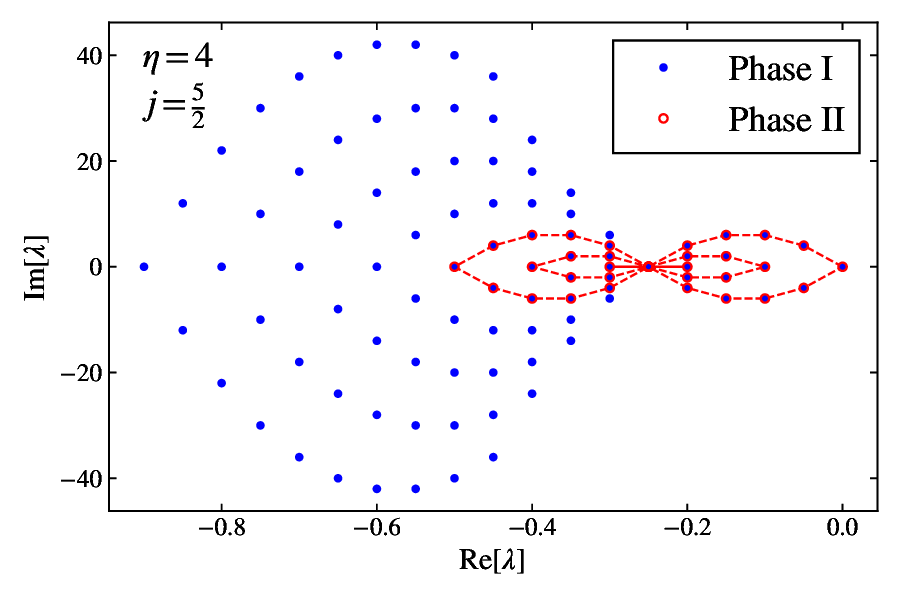}
    \caption{Scatterplot of the Kerr oscillator Liouvillian spectrum for $\omega / K = \eta = 4$ showing the separation of eigenvalues into two phases, Phase I (blue) and Phase II (red).}
    \label{fig:fig3}
\end{figure}

The eigenvalues of the Hamiltonian of the Kerr oscillator have a two phase
structure \cite{iac3}. In Phase I states are singly degenerate, while in
Phase II they are doubly degenerate. It has been found that Phase II has,
for integer values of $\eta$, a remarkable quasi-spin $su(2)$ symmetry,
with states characterized by quasi-spin representations $\left\vert j,m_{j}\right\rangle $,
with $j = \frac{\eta+1}{2}$ and energies counted from the lowest state given
by $E-E_{\rm min}=m_{j}^{2}$ for integer $j$ and $E-E_{\rm min}=m_{j}^{2} - \frac{1}{4}$
for half integer $j$ \cite{iac3}. This quasi-spin symmetry $su(2)$ is not the same as the
symmetry of the one-dimensional harmonic oscillator discussed in subsection~\ref{spectral1d},
and it represents a major novel finding. Details of the derivation of the
quasi-spin symmetry $su(2)$, its construction using two boson operators,
and its role in determining the eigenvalues of the Hamiltonian $\hat{H}$ are given
in \cite{iac3} and~\ref{appendixa}. It is a remarkable result of the present article that both
the two phase structure and the $su(2)$ quasi-spin symmetry appear in the
eigenvalues of the Liouville superoperator. The eigenvalues of $\mathcal{L}$ for
Phase I have the same structure of the anharmonic oscillator, $\omega=-1,K=1$
(figure \ref{fig:fig2}a). The eigenvalues of $\mathcal{L}$ for Phase II have a
characteristic double-ellipsoidal structure which is different for $\eta=\mathrm{even}$
and $\eta=\mathrm{odd}$, and a highly degenerate accumulation point. The two
structures are interpenetrating, as shown for $\eta = \omega / K = 4$ in figure \ref{fig:fig3},
where the two structures are color coded, Phase I in blue and Phase II in red.
Note that the accumulation point at the center of the double-ellipsoidal structure of Phase II is six-fold degenerate.

\section{Quasi-spin symmetry of the Kerr oscillator Liouvillian} \label{kpoquasispin}

\begin{figure}[ht!]
    \centering
    \includegraphics[width=0.4\textwidth]{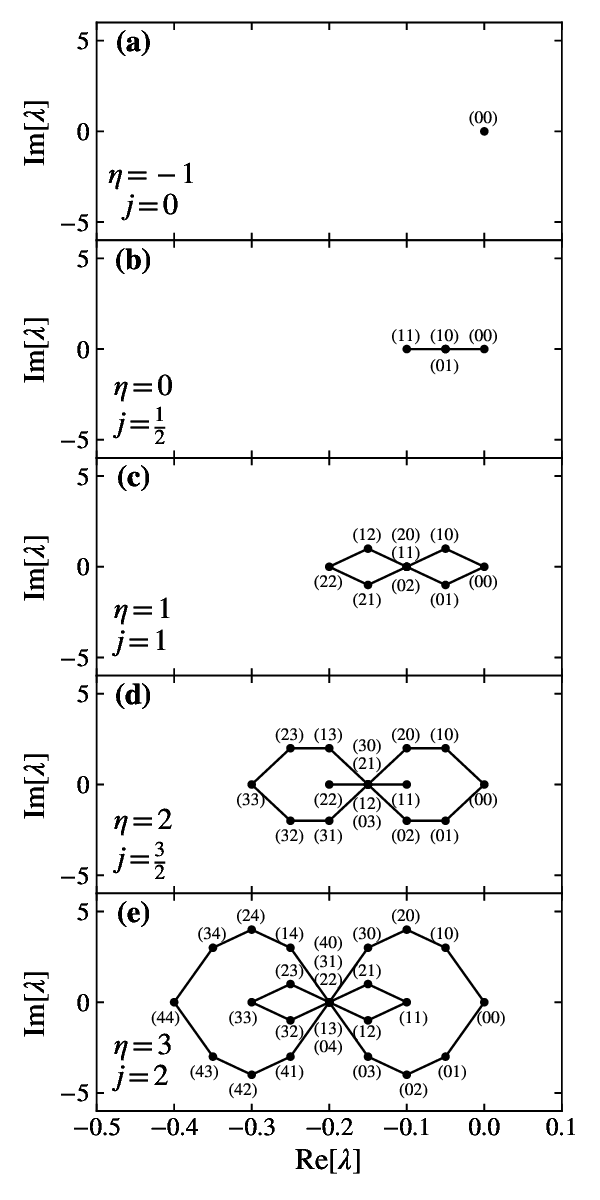}
    \caption{Eigenvalues of the Liouvillian superoperator of the Kerr oscillator for irreducible
representations $\left\vert j,m_{j}\right\rangle$
of $su(2)_{j}$ with $j=0,1/2,1,3/2,2$.
Points are labeled by $(n,m) = \left(j-m_{j}, j-m_{j}^{\prime} \right)$,
(\ref{eq:kerrlioueigs}). Note the accumulation point at the center of the double-ellispoidal structure.}
    \label{fig:fig4}
\end{figure}

The $su(2)$ quasi-spin symmetry of the Kerr oscillator Liouvillian appears as a consequence of the
Hamiltonian symmetry. As discussed in~\ref{appendixa}, the Hamiltonian projected onto the Phase II subspace of the
Kerr oscillator can be written in quasi-spin notation as
\begin{equation}
\label{eq:quasispinham}
\frac{\hat{H}}{K} = \hat{j}_{z}^{2} - \frac{1}{8} \left( 1 - (-1)^{2j} \right),
\end{equation}
acting on states $\left\vert j,m_{j}\right\rangle$ where
$\hat{j}_{z} \left\vert j,m_{j}\right\rangle = m_{j} \left\vert j,m_{j}\right\rangle$. Similarly making use of
the two-boson construction of representations of $su(2)$ described in~\ref{appendixa}, one can
characterize the eigenvalues of the Liouvillian for Phase II in terms of
an $su(2)$ representation $\left\vert j,m_{j}\right\rangle$ $\left(j = \frac{\eta^{\prime}}{2} = \frac{N}{2} \right)$
and its dual representation $j^{\ast}$ (which for $su(2)$
is isomorphic to the original representation $j$ \cite{iac5}).
Eigenvalues, as shown in figure \ref{fig:fig4}, can then be labeled by $( m_{j}, m_{j}^{\prime} )$
and (\ref{eq:kerrlioueigs}) can be rewritten as
\begin{equation}
\label{eq:kerrlioueigssu2}
\lambda _{j,m_{j},m_{j}^{\prime}}=-\rmi\left(m_{j}^{2} -m_{j}^{\prime 2} \right) -\kappa \left( j - \frac{m_{j} + m_{j}^{\prime}}{2} \right)
\end{equation}
where $n = (j-m_{j})$ and $m = (j-m_{j}^{\prime})$. Associated eigenmatrices are denoted $\hat{\rho}_{j,m_{j},m_{j}^{\prime}}$.
Furthermore, we can express the Liouvillian $\mathcal{L}$ projected onto the Phase II subspace in terms of quasi-spin operators,
\begin{equation}
\label{eq:kerrliousu2}
\mathcal{L} \hat{\rho} = -\rmi \left[ \hat{j}_{z}^{2} - \frac{1}{8} \left( 1 - (-1)^{2j} \right), \hat{\rho} \right] -
\frac{\kappa}{2} \left( (j - \hat{j}_{z} ) \hat{\rho} + \hat{\rho} (j - \hat{j}_{z} ) \right)
\end{equation}
The quasi-spin dynamic symmetry of the Liouvillian is evident here, as
$\mathcal{L}$ consists exclusively of invariant operators of $su(2)_{j}$ acting on $\hat{\rho}$ from the left,
and invariant operators of the dual $su(2)_{j^{\ast}}$ acting on $\hat{\rho}$ from the right.
It is important to clarify that eigenmatrices of  $\mathcal{L}$ in the Phase II subspace,
$\hat{\rho}_{j,m_{j},m_{j}^{\prime}}$,
are not the same as outer products of Hamiltonian eigenstates,
$\left\vert j,m_{j}\right\rangle \left\langle j,m_{j}^{\prime} \right\vert$, as discussed in subsection~\ref{spectral1d}.
Rather, the notation $( m_{j}, m_{j}^{\prime} )$ simply labels Liouvillian
eigenmatrices with $su(2)_{j}$ quantum numbers.

From a mathematical point of view, it is interesting to display the
eigenvalues of the Liouville\ superoperator for irreducible representations
of $su(2)_{j}$,
as shown in figure \ref{fig:fig4}. The symmetries of the Liouvillian are particularly evident in this figure.
In addition to the reflection on the $\mathrm{Re}[\lambda]$ axis due to the definition of the Liouvillian
(\ref{eq:liouvillesplit}), there is a symmetry of reflection on the $\mathrm{Im}[\lambda]$ axis at
$\mathrm{Re}[\lambda] = -\kappa j$ ($\kappa = 0.1$ in the figure). This symmetry is a
consequence of the degeneracy of the Hamiltonian, and allows points to the left and to the right of
the accumulation point $\mathrm{Re}[\lambda] = -\kappa j$ to be classified separately in conjugate pairs
of $su(2)$ representations. To be specific, points
to the right of the accumulation point can be classified as representations of $su(2)$,
$\left\vert J, M \right\rangle$ with $0 \le J \le j$.
Points to the left can be classified in a similar way as $\left\vert \bar{J}, \bar{M} \right\rangle$, where
$\left\vert \bar{J}, \bar{M} \right\rangle$ are conjugate representations of $su(2)$ obtained from
$\left\vert J, M \right\rangle$ by reflection.
These $su(2)$ representations are obtained from the
$( m_{j}, m_{j}^{\prime} )$ or the $(n,m)$ classification by
\begin{eqnarray}
\label{eq:su2classification}
J &=& j - \frac{m_{j} + m_{j}^{\prime}}{2} = \frac{n + m}{2}  \nonumber \\
M &=& \frac{m_{j}^{\prime} - m_{j}}{2} = \frac{n - m}{2}  \nonumber \\
\bar{J} &=& j + \frac{m_{j} + m_{j}^{\prime}}{2} = (\eta + 1) - \frac{n + m}{2}  \nonumber \\
\bar{M} &=& \frac{m_{j} - m_{j}^{\prime}}{2} = -\frac{n - m}{2}
\end{eqnarray}
As an example of this classification, the accumulation point at the center of the double-ellipsoidal structure in figure \ref{fig:fig4}e,
$\eta = 3, j = 2$, has $(n,m) = (40), (31), (22), (13), (04)$, forming a representation
\cite{iac5} $J = 2$ and $M = \pm 2, \pm 1, 0$. In this figure
starting from the right and moving to the left, one has $su(2)$ representations grouped in vertical bands
$J = 0,\frac{1}{2}, 1, \frac{3}{2}$; $J = \bar{J} = 2$; $\bar{J} =  \frac{3}{2}, 1, \frac{1}{2}, 0$.

The spectrum of the Liouvillian in Phase II can also be expressed in terms of $\left\vert J, M \right\rangle$ and
$\left\vert \bar{J}, \bar{M} \right\rangle$ representations, where, denoting eigenmatrices $\hat{\rho}_{j,J,M}$ and
$\hat{\rho}_{j,\bar{J},\bar{M}}$, one obtains
\begin{eqnarray}
\label{eq:su2conjreps}
\mathcal{L} \hat{\rho}_{j,J,M} &= \left[ 4 \rmi \left(j - J \right) M - \kappa J \right] \hat{\rho}_{j,J,M} & \qquad
0 \leq J \leq j \nonumber \\
\mathcal{L} \hat{\rho}_{j,\bar{J},\bar{M}} &= \left[ 4 \rmi \left(j - \bar{J} \right) \bar{M} - \kappa \left(2j - \bar{J} \right) \right]
\hat{\rho}_{j,\bar{J},\bar{M}} & \qquad 0 \leq \bar{J} \leq j
\end{eqnarray}
Note that $J = j$ and $\bar{J} = j$ denote the same representation, so one must be careful not to overcount here.
All three different forms, $(n,m)$, $( m_{j}, m_{j}^{\prime} )$, $( J,M ) ( \bar{J},\bar{M} )$,
are equivalent and can be obtained from each other by the relation given above.

Particularly interesting is the Liouvillian of a spinor, $j = \frac{1}{2}$, consisting of a line, since
the two Hamiltonian eigenstates $n=0$ and $n=1$ are degenerate and therefore
$\mathrm{Im}[\lambda] = 0$. These degenerate spinor states may be used to form a basis for a qubit.

We note that numerical diagonalization confirms the analytic formulas
derived above, and that the $su(2)$ symmetry of the Liouville superoperator at
integer values of the parameter $\eta $ has key implications for the stabilization of long-lived states,
as will shown be in section~\ref{spectraltheorysqueezed}.

\section{Spectral theory of one-dimensional squeezed oscillators} \label{spectraltheorysqueezed}

One-dimensional squeezed oscillators have Hamiltonians of the form
\begin{equation}
\label{eq:1dsqueezedosc}
\hat{H}=\sum_{k}\omega _{k}\left( \hat{a}^{\dag }\hat{a}\right)
^{k}+\sum_{k_{\rm s}}\varepsilon _{k_{\rm s}}\left( \hat{a}^{\dag k_{\rm s}}+\hat{a}^{k_{\rm s}}\right)
\end{equation}
where $k_{\rm s}$ denotes the order of the squeezing, and zero-temperature linear
dissipators $\mathcal{D}[\hat{a}]\hat{\rho}(t)$ given in (\ref{eq:kerrlineardissipator}). The eigenvalues
of the Liouville superoperator for squeezed oscillators cannot be solved analytically even at
zero-temperature, except when the oscillator is linear, $k=1$, and the squeezing is (at most)
quadratic, $k_{\rm s} \leq 2$ \cite{prosen}, in which case the spectrum can be obtained analytically
in the dynamically stable regime. $\hat{H}$ still has a $h(2)$ spectrum generating algebra
but no longer has $u(1)$ symmetry. Therefore, its eigenfunctions are of the form
\begin{equation}
\label{eq:squeezedoscket}
\left\vert \psi _{\alpha }\right\rangle =\sum_{n=0}^{\infty }c_{n}^{(\alpha
)}\left\vert n\right\rangle
\end{equation}
and we can no longer use the $u(1)$ symmetry to diagonalize
$\mathcal{L}$ as was done in the previous section.
In this case, the eigenvalues
must be evaluated numerically in a truncated Hilbert space of dimension $N_{\rm Fock}=N+1$.
The dimension of the Hilbert space plays here an important
role especially for large values of the parameters $\varepsilon _{k_{\rm s}}$,
since the truncated space is invariant under the action of
$\hat{a}^{\dag }\hat{a}$ and $\hat{a}$, but not of $\hat{a}^{\dag }$. Thus, careful attention
must be given to $N_{\rm Fock}$ to ensure convergence of eigenvalues and properties of interest.

In this article, we consider squeezed Kerr oscillators with linear dissipation,
the class of models with Hamiltonian $\hat{H}$ and Liouvillian
$\mathcal{L} = \mathcal{L}^{(0)} + \mathcal{L}^{(1)}$ given by 
\begin{eqnarray}
\label{eq:squeezedkerrliou}
\hat{H} &=&-\omega \hat{a}^{\dag }\hat{a}+K\hat{a}^{\dag 2}\hat{a}^{2}-\varepsilon _{2}\left( \hat{a}^{\dag 2}+\hat{a}^{2}\right)  \nonumber \\
\mathcal{L}^{(1)}\hat{\rho} &=&\kappa \left( \hat{a}\hat{\rho}\hat{a}^{\dag}-\frac{1}{2}(\hat{a}^{\dag }\hat{a}\hat{\rho}+\hat{\rho}\hat{a}^{\dag }\hat{a}\right) .
\end{eqnarray}
These models have been proposed as devices for robust quantum computation \cite{goto2}
and squeezing can be implemented experimentally with
superconducting circuits \cite{frattini,venkatraman1,venkatraman2}.

\subsection{Evolution of the Liouvillian eigenvalues as a function of $\varepsilon _{2}/K$} \label{lioueigenvals}

We have studied numerically the eigenvalues of the Liouville superoperator
for the harmonic oscillator and the squeezed Kerr oscillator as a function
of the parameter $\varepsilon _{2}$. In all scatterplots shown below we have
used $\kappa =0.1$.

\subsubsection{Squeezed Harmonic oscillator.} \label{squeezedqhoeigs}

\begin{figure}[ht!]
    \centering
    \includegraphics[width=0.7\textwidth]{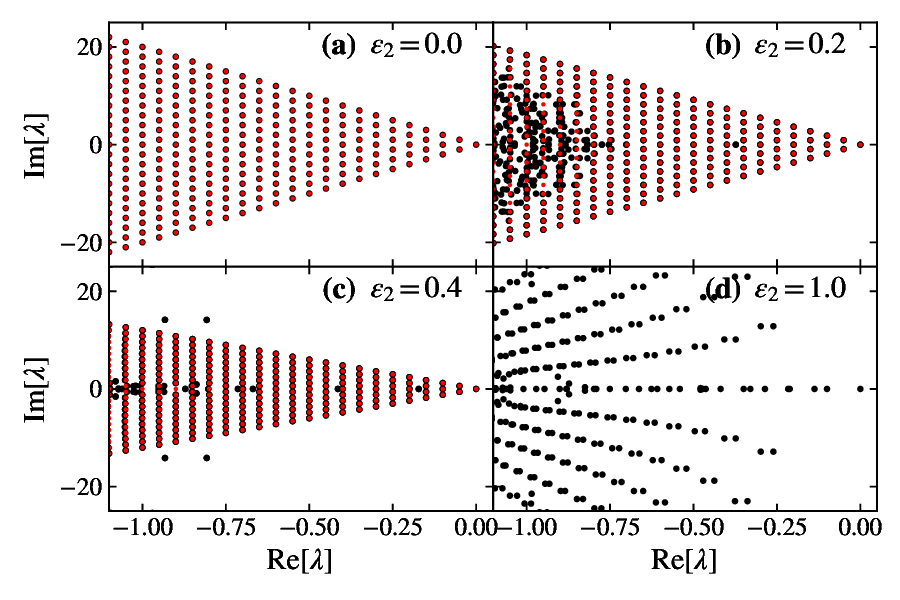}
    \caption{Scatterplot of the squeezed harmonic oscillator ($\omega =-1,K=0$) Liouvillian spectrum for $\varepsilon_{2}=0,0.2,0.4,1.0$, and $N_{\rm Fock} = 120$. In panels (a)-(c), red points are given by the analytic formula (\ref{eq:squeezedqholiou}).}
    \label{fig:fig5}
\end{figure}

The Hamiltonian for this case is
\begin{equation}
\label{eq:squeezedqhoham}
\hat{H} =-\omega \hat{a}^{\dag }\hat{a} - \varepsilon _{2}\left( \hat{a}^{\dag 2}+\hat{a}^{2}\right) =
-\omega \left[ \hat{a}^{\dag }\hat{a} + z \left( \hat{a}^{\dag 2}+\hat{a}^{2}\right) \right].
\end{equation}
This Hamiltonian has a second order quantum phase transition (QPT) at
$z = \varepsilon _{2} / | \omega | = 0.5$ \cite{vanroosmalen}. Its Liouvillian spectrum is shown in
figure \ref{fig:fig5} as a function of squeezing $\varepsilon _{2}$ for $N_{\rm Fock} = 120$. A large value
of $N_{\rm Fock}$ is used here to ensure convergence. For $z = \varepsilon _{2} / | \omega | \leq 0.5$,
an explicit formula for the eigenvalues can be derived due to the fact that the Liouvillian is quadratic, via
third quantization methods \cite{viola1,buca,prosen,barthel,viola2},
\begin{equation}
\label{eq:squeezedqholiou}
\lambda _{n,m}=-\rmi \sqrt{\omega^2 - 4 \left(\varepsilon _{2}\right)^2} \left( n_1 - n_2 \right) -\frac{\kappa }{2}(n_1+n_2)
\end{equation}
with $n_1,n_2$ non-negative integers. The spectrum of the squeezed harmonic oscillator is similar to
that of the unsqueezed harmonic oscillator (\ref{eq:harmonicoscliouvillian}) but with a renormalized frequency
$\sqrt{\omega^2 - 4 \left(\varepsilon _{2}\right)^2}$.

Numerical diagonalization for $\varepsilon _{2} / | \omega | \leq 0.5$ agrees with the formula
(\ref{eq:squeezedqholiou}) up to some anomalous points due to truncation of the
computational Hilbert space, $N_{\rm Fock}$.
For  $\varepsilon _{2} / | \omega | > 0.5$, the situation is complicated. The QPT that
for the Hamiltonian occurs at $z = \varepsilon _{2} / | \omega | = 0.5$ is modified when
the Liouvillian is considered. A new phase occurs in the region
$ \frac{| \omega | }{2} < \varepsilon_{2} < \frac{1}{2} \sqrt{\omega^2 + \kappa^2 / 4}$. The situation is
similar to that described in section \ref{liouqpts} for the Kerr oscillator. Since the aim
of this paper is to study the Kerr oscillator, we do not discuss it further. We only mention
here that no explicit formula is known for the eigenvalues of the Liouvillian for
$\varepsilon_{2} > \frac{1}{2} \sqrt{\omega^2 + \kappa^2 / 4}$. An approximate formula
can be derived as in the following subsection.

\subsubsection{Squeezed Kerr oscillator.} \label{squeezedkpoeigs}

\begin{figure}[ht!]
    \centering
    \includegraphics[width=0.9\textwidth]{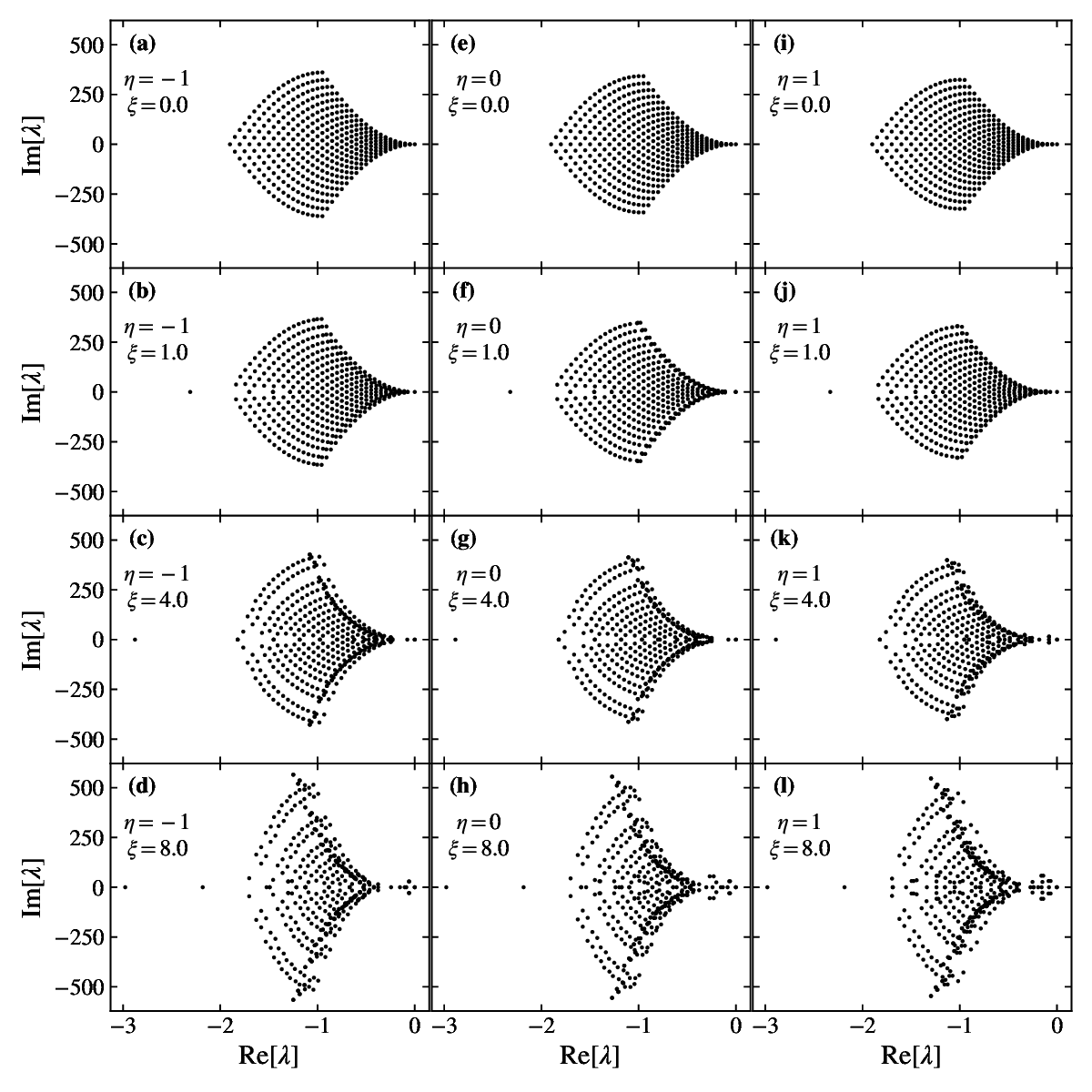}
    \caption{Scatterplot of the squeezed Kerr oscillator for $\eta = \omega / K =-1,0,+1$, $\xi = \varepsilon _{2} / K=0.0,1.0,4.0,8.0$, and $N_{\rm Fock} = 20$.}
    \label{fig:fig6}
\end{figure}

We have investigated the Liouvillian spectrum of the
squeezed Kerr oscillator for $\omega =-1,0,1$, $K=1$. The dimensionless
Hamiltonian for these cases is
\begin{equation}
\label{eq:dimlesssqueezedkerr}
\frac{\hat{H}}{K}=-\eta \hat{n}+\hat{n}(\hat{n}-1)-\xi \hat{P}_{2}
\end{equation}
where $\hat{n}=\hat{a}^{\dag }\hat{a}$ is the number operator,
$\hat{P}_{2}=\left( \hat{a}^{\dag 2}+\hat{a}^{2}\right) $ is the pairing operator of
order 2 and $\eta =\omega / K, \xi =\varepsilon _{2} / K$. Scatterplots
of eigenvalues for these three cases, $\eta =-1,0,+1$, needed in the study of
the quantum phase transitions to be discussed in the next section, are shown
in figure \ref{fig:fig6}. It is important to note that numerical results converge for
much smaller $N_{\rm Fock}$ than for the harmonic oscillator, due to the presence of
the Kerr nonlinearity in the Hamiltonian, $\hat{n}^2$.

As $\xi$ increases, one can observe the formation of structures similar to those of figure \ref{fig:fig4}. These structures are particularly evident for $\xi = 4.0$ and $\xi = 8.0$. Their actual form is similar but not identical to that of figure \ref{fig:fig4}, since here the energy of the states belonging to the representation $\left\vert j^{\prime },m^{\prime }\right\rangle $ of $su(2)$ are $E_{m^{\prime}_{j^{\prime}}} = \vert m^{\prime}_{j^{\prime}} \vert$ \cite{iac3} and not $E_{m^{\prime}_{j^{\prime}}} = {m^{\prime}_{j^{\prime}}}^2$. We denote the representations of $su(2)$ here as $\left\vert j^{\prime },m^{\prime }\right\rangle$ since they are different from the representations $\left\vert j, m_{j} \right\rangle$ discussed previously.

\begin{figure}[ht!]
    \centering
    \includegraphics[width=0.9\textwidth]{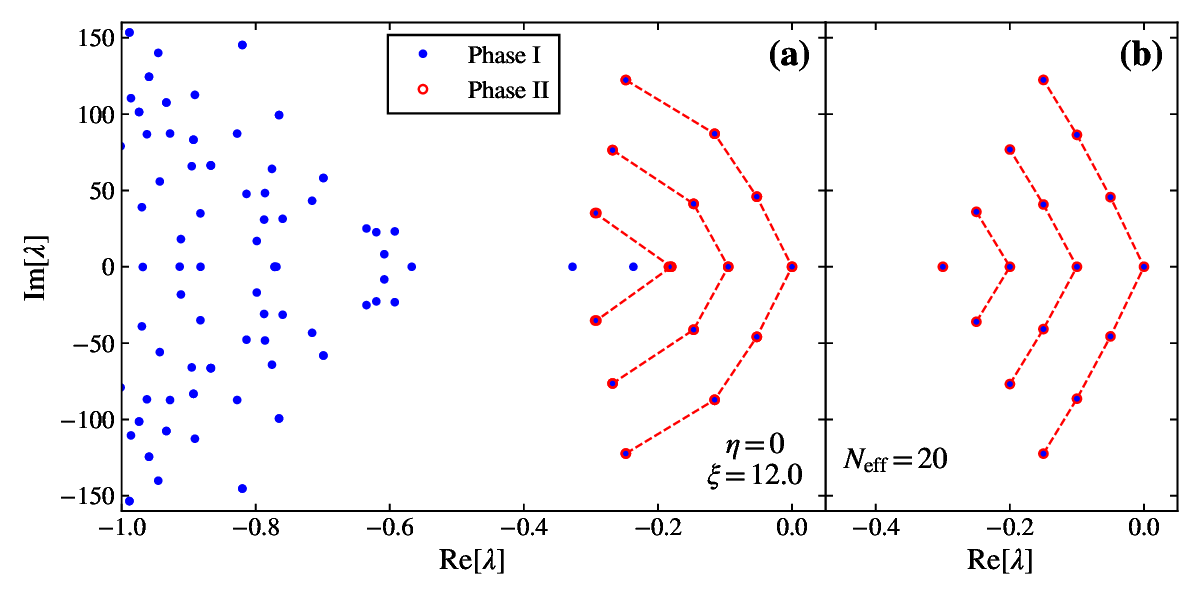}
    \caption{(a) Portion $\left\vert \mathrm{Re}\left[ \lambda \right] \right\vert \leq 1$ of the scatterplot of the squeezed Kerr oscillator for $\eta = \omega / K = 0$ and $\xi = \varepsilon_{2} / K = 12.0$. Doubly degenerate points belonging to Phase II are colored in red. Here $N=80 = 2 N_{\rm conv}$ to ensure convergence of Phase I and Phase II eigenvalues for $\xi=12.0$. (b) Scatterplot of the doubly degenerate anharmonic oscillator of (\ref{eq:oscnu}) with $\nu = 0,1,2,3$, $\pi = \pm$, and $N_{\rm eff} = 20$. All points are doubly degenerate and colored in red.}
    \label{fig:fig7}
\end{figure}

This form is illustrated in figure \ref{fig:fig7}a, where the portion $\left\vert \mathrm{Re}\left[ \lambda \right] \right\vert \leq 1$ of the scatterplot for $\eta = \omega / K = 0$, $\xi = \varepsilon_{2} / K = 12.0$ is shown, with doubly degenerate points marked in red. The two phases are clearly separated here. The structure of Phase II is a perturbed form of a doubly degenerate anharmonic oscillator with energies given approximately by equation (52) of \cite{iac3},
\begin{equation}
\label{eq:anharmonicenergy}
E_{\nu}^{\pi} = 4 \xi \nu \left( 1 - \frac{\nu}{N_{\rm eff}} \right)
\end{equation}
where $\nu^{\pi}$ $(\nu = 0,1,... ; \pi = \pm)$ denotes the quantum states, and $N_{\rm eff} = N_{\rm conv} / 2$ where $N_{\rm conv}$ is the value $N_{\rm Fock} - 1$ of states required in the numerical calculation for convergence of eigenvalues in Phase II. The Liouvillian spectrum of the oscillator is
\begin{equation}
\label{eq:oscnu}
\lambda _{\nu,\nu^{\prime}}=-\rmi \left( E_{\nu} - E_{\nu^{\prime}} \right) -\frac{\kappa }{2} \left(\nu + \nu^{\prime} \right)
\end{equation}
shown in figure \ref{fig:fig7}b. For the first three branches, the imaginary parts of Phase II eigenvalues, $\mathrm{Im}\left[ \lambda \right]$, follow closely (\ref{eq:oscnu}). However, the real parts, $\mathrm{Re}\left[ \lambda \right]$, are highly perturbed. Moreover, the fourth branch consisting of a  single doubly degenerate point does not appear in panel (a) due to mixing with the singly degenerate states of Phase I.

\begin{figure}[ht!]
    \centering
    \includegraphics[width=0.8\textwidth]{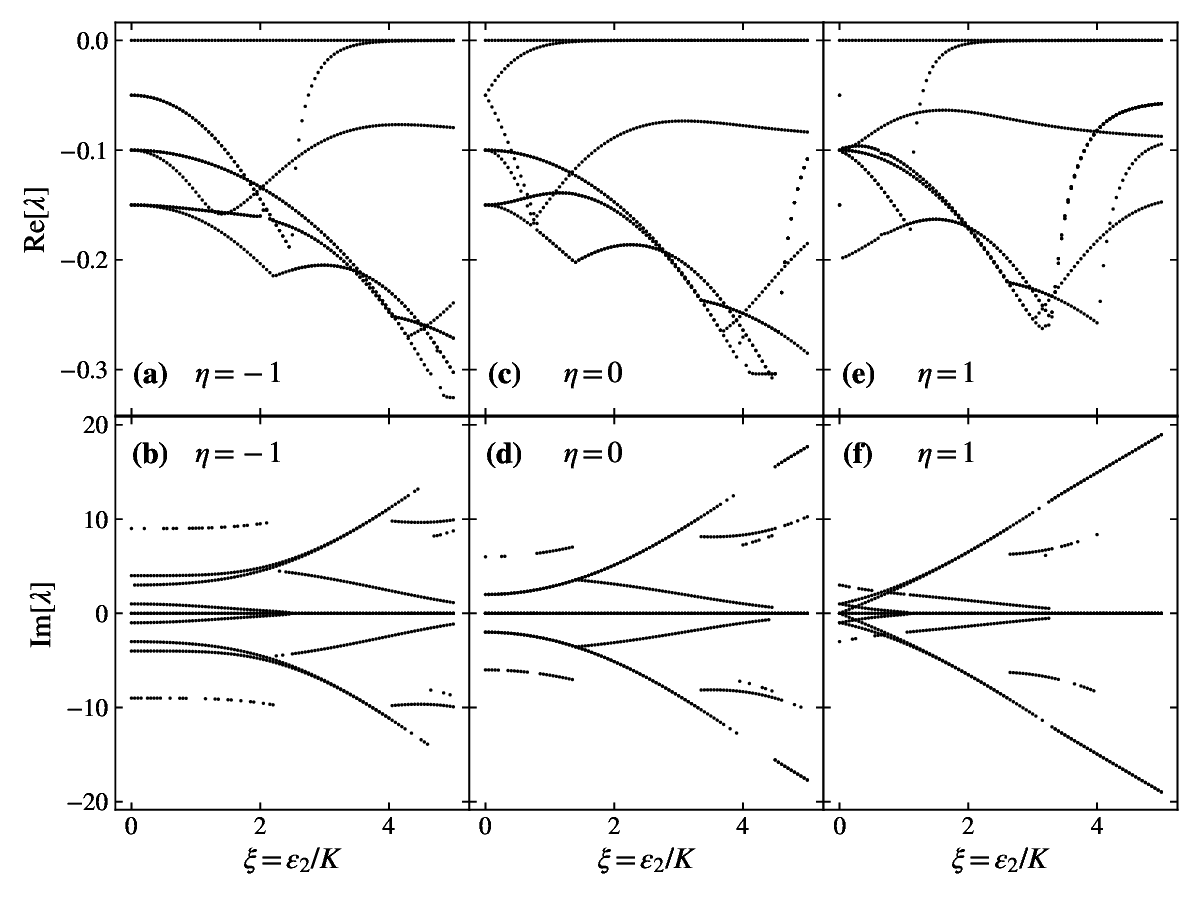}
    \caption{The lowest eigenvalues $\lambda _{i}(i=1,...,9)$ of the Liouville
superoperator for the Kerr oscillator for $\eta =\omega /K=-1,0,+1$ as a
function of $\xi =\varepsilon _{2} / K$. $N_{\rm Fock}=80$ in all calculations.}
    \label{fig:fig8}
\end{figure}

\begin{figure}[ht!]
    \centering
    \includegraphics[width=0.4\textwidth]{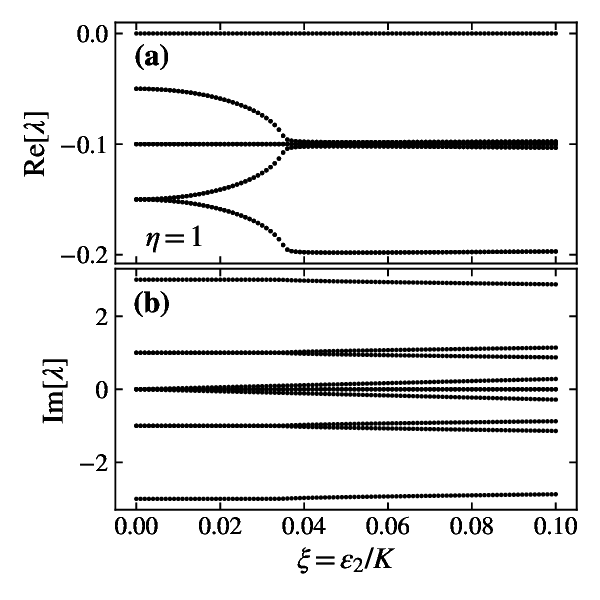}
    \caption{The small $\xi $ behavior of $\mathrm{Re}[\lambda _{i}]$ for the Kerr oscillator for $\eta =+1$.}
    \label{fig:fig9}
\end{figure}

In order to clarify further the situation, we consider next the evolution of
the eigenvalues of the Liouville superoperator for the squeezed Kerr
oscillator as a function of the parameter $\xi =\varepsilon _{2}/K$
separated into real and imaginary part. In figure \ref{fig:fig8}, we show $\mathrm{Re}[\lambda
_{i}]$ and $\mathrm{Im}[\lambda _{i}]$ for the first nine eigenvalues. From this figure one can see that $\mathrm{Re}[\lambda _{1}]$ for $\eta =1$
has a discontinuity around $\xi =0$. To clarify this behavior we show in
figure \ref{fig:fig9} a close up of $\mathrm{Re}[\lambda _{i}]$ for small $\xi $. The
discontinuity is related to a 1st order QPT which will be discussed in the
following section.

\begin{figure}[ht!]
    \centering
    \includegraphics[width=0.7\textwidth]{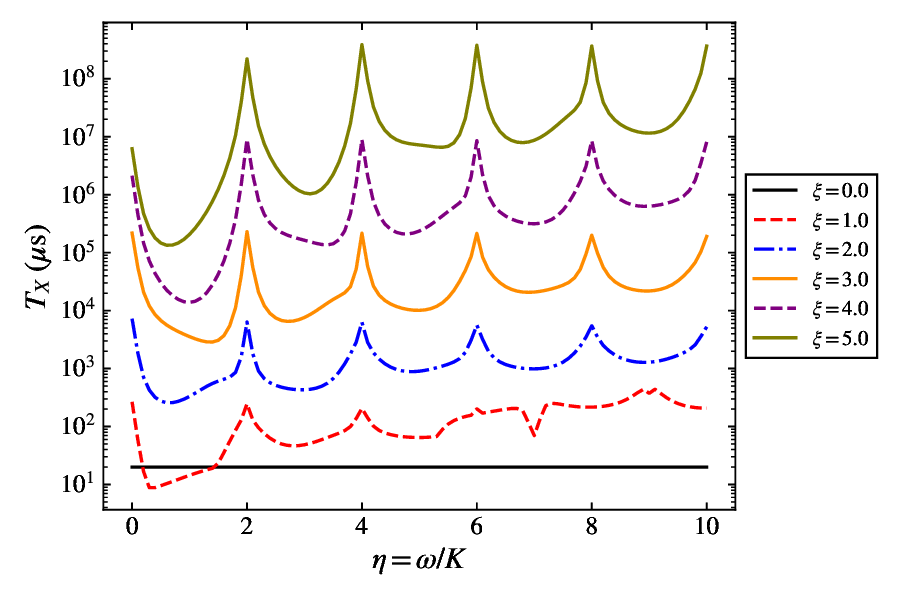}
    \caption{The relaxation time $T_{X}$ of the squeezed Kerr oscillator for $0\leq \eta \leq 10$, $0\leq \xi \leq 5$, and $\kappa = 0.1$ $\mu$s$^{-1}$. Here $N_{\rm Fock}=80$ is used.}
    \label{fig:fig10}
\end{figure}

From the lowest nonzero eigenvalue $\lambda _{1}$ we can calculate the relaxation time
\begin{equation}
\label{eq:relaxtime}
T_{X}=-\frac{1}{\mathrm{Re}[\lambda _{1}]}
\end{equation}
shown in figure \ref{fig:fig10} for some
values of $\xi $ and as a function of $\eta $. We see that for large $\xi$
a characteristic behavior emerges, that is for even values of the parameter $\eta$
the relaxation time is larger than for odd values. The peak at $\eta=2$
already appears for $\xi =0.5$, while those at $\eta =4,6,8,10$ appear
at $\xi =\eta /4=1,1.5,2.0,2.5$.
These peaks at integer $\eta$ can be attributed to the quasi-spin symmetry of the
Liouvillian maintained at nonzero $\xi$, discussed previously. Maxima in $T_{X}$ occur at even
$\eta$ due to degeneracies of Hamiltonian eigenvalues maintained for finite $\xi$ and smaller
$T_{X}$ occur at odd $\eta$ due to avoided level crossings.
These Hamiltonian properties are further discussed in \cite{iac3}.
This result is of particular importance
for quantum computing, since it shows that by appropriately tuning the
parameters one can devise systems with large relaxation time.

\section{QPT and ESQPT in open systems} \label{qpts}

\subsection{QPT and ESQPT of the Hamiltonian operator} \label{hamqpts}

Quantum Phase Transitions (QPT) \cite{sachdev} and associated Excited State
Quantum Phase Transitions (ESQPT) \cite{caprio,cejnar1,cejnar2}
of Hamiltonian operators have been extensively studied in a
variety of systems \cite{carr}, especially for algebraic models with
structure $g\equiv su(n)$ \cite{caprio}. For the Hamiltonian operator the
consequences of a QPT are: (i) the ground state energy $E_{0}$ is a
non-analytic function of the control parameter at $\xi =\xi _{\rm c}$; (ii) the
ground state wave function properties, expressed via ``order parameters'',
i.e. the expectation value of some suitable chosen operator
$\left\langle \hat{o}\right\rangle $, are non-analytic at $\xi =\xi _{\rm c}$;
(iii) the energy gap $\Delta_{E_{1}} = E_{1} - E_{0}$
between the ground state and the first excited state vanishes at $\xi =\xi _{\rm c}$.
For finite systems with $N$ constituents, the defining
characteristic of a QPT is not the presence of a true singularity but rather
well-defined scaling properties of the relevant quantities towards their
singular $N\rightarrow \infty $ limits, called the thermodynamic limit. QPTs are
called 0th, 1st, 2nd,... order if the discontinuity occurs in the ground
state energy, $E_{0}$, or in the first, second,... derivative $\partial
E_{0}/\partial \xi ,\partial ^{2}E_{0}/\partial \xi ^{2},...$ (Ehrenfest
classification). In the associated order parameter $\left\langle \hat{o}\right\rangle $,
the discontinuities occur for first order in $\left\langle \hat{o}\right\rangle $,
second order in $\partial \left\langle \hat{o}\right\rangle /\partial \xi$, ...

The Hamiltonian operator $\hat{H}$ of (\ref{eq:dimlesssqueezedkerr}) has two control parameters, $\eta $
and $\xi $. It has already been found that, for $\eta =0$ and $\xi =0$,
there is a 2nd order QPT and, as a function of $\xi $, an associated ESQPT 
\cite{prado,chavez}. The phase structure of the Hamiltonian operator
of the squeezed Kerr oscillator is however more complicated than a single
2nd order transition. 

\begin{figure}[ht!]
    \centering
    \includegraphics[width=0.65\textwidth]{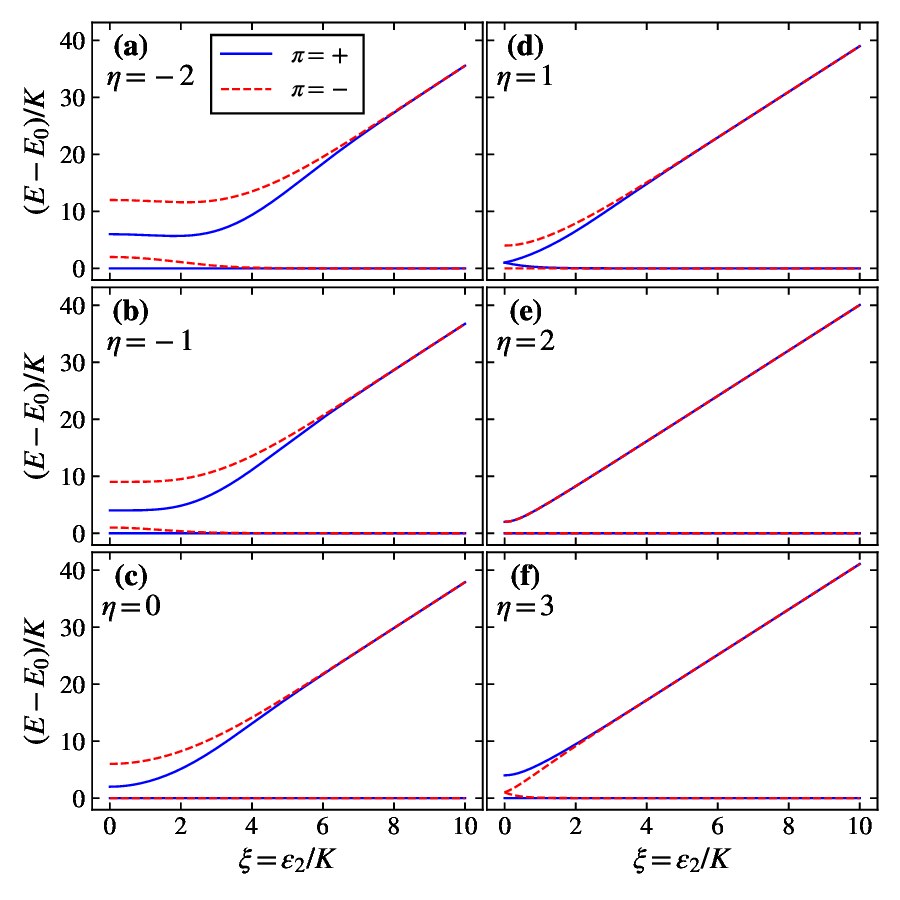}
    \caption{The energy levels of the squeezed Kerr oscillator as a function of $\xi $
for $-2\leq \eta \leq +3$. Levels are labeled by parity $\pi =\pm $.}
    \label{fig:fig11}
\end{figure}

To illustrate this point, we show in figure \ref{fig:fig11} a plot
of the lowest four energy levels as a function of $\xi $ for $-2\leq \eta
\leq +3$. One can see here clearly the occurrence of property (iii) (vanishing of the Hamiltonian gap), indicating a transition at values $\xi_{\rm k}$ given by $\xi_{\rm k} = -2 \eta$ for $\eta = -2,-1,0$, but a more complex structure for $\eta >0$. The value $\xi_{\rm k}$ at which $\Delta_{E_{1}} = E_{1} - E_{0} = 0$ is also called a ``kissing point'', hence the index $\rm k$ given to $\xi$.

\begin{figure}[ht!]
    \centering
    \includegraphics[width=0.65\textwidth]{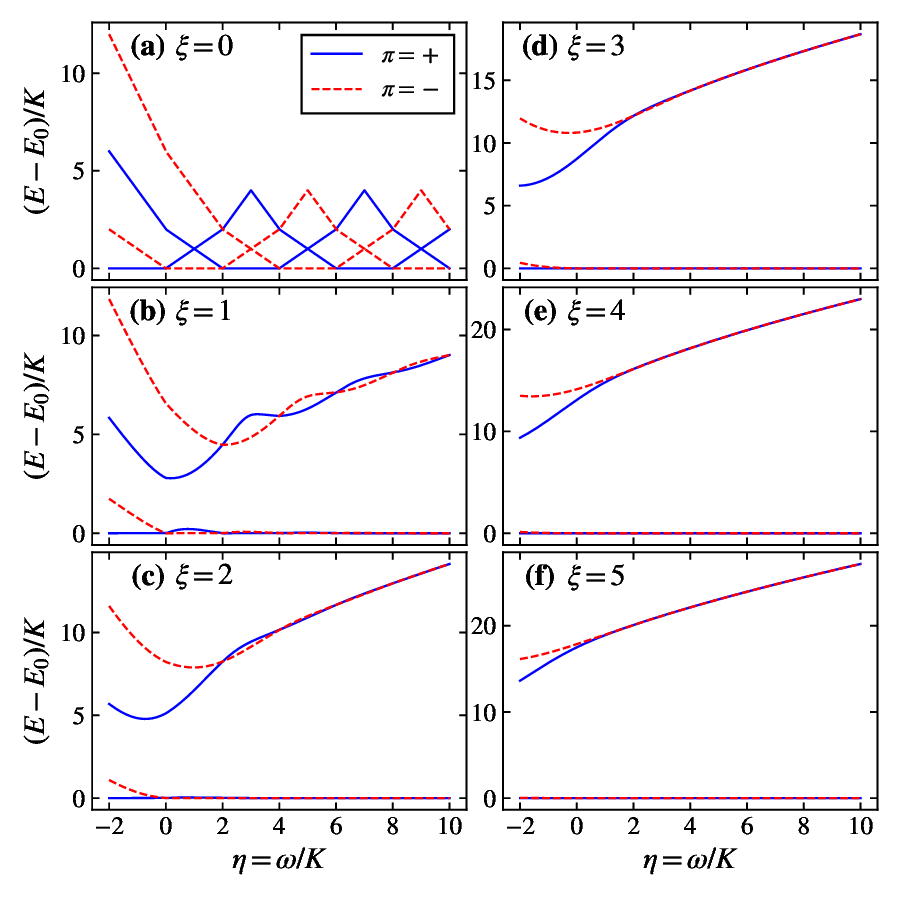}
    \caption{The energy levels of the squeezed Kerr oscillator as a function of $\eta $
for $\xi =0,1,...,5$. Levels are labeled by parity $\pi =\pm $.}
    \label{fig:fig12}
\end{figure}

In figure \ref{fig:fig12}, we show a plot of the lowest four energy levels as a function of $\eta $
for $0\leq \xi \leq 5$. The behavior of the energy levels as a function of $\eta$ for fixed $\xi$
is rather complex. At $\xi =0$, it is dictated by the quasi-spin symmetry $su(2)$,
as shown in~\ref{appendixa}. It consists in a succession of crossing of
levels of opposite parity at integer values of $\eta =0,1,2,$... At $\xi >0$
the level crossings persist, as seen for example at $\xi =1$ and $\xi =2$,
but for large $\eta $ and/or large $\xi $ levels become doubly degenerate. This
situation, with the occurrence of three phases I, II, III was described in
detail in \cite{iac3}. Its description in terms of phase transitions is rather
complex and it requires the study of the classical limit of the
algebraic Hamiltonian. We therefore defer it to a later publication.

\begin{figure}[ht!]
    \centering
    \includegraphics[width=0.7\textwidth]{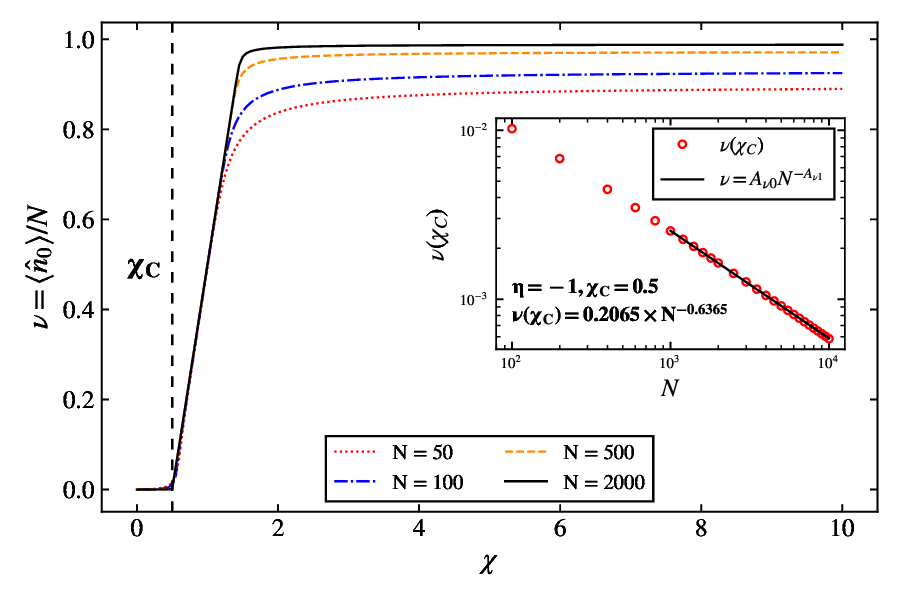}
    \caption{Order parameter for $\eta =-1$ as a function of $\chi $ for
different values of $N$. The critical value is $\chi_{\rm c}=0.5$, shown by a dashed vertical line. The scaling behavior of the order parameter at $\chi_{\rm c}$ is plotted in the insert.}
    \label{fig:fig13}
\end{figure}

For 2nd order transitions of Hamiltonian operators, it is possible to
find scaling properties and study the so-called thermodynamic limit $N\rightarrow \infty $,
as, for example, in \cite{bernal}. To this end, it is
convenient to construct the scaled Hamiltonian 
\begin{equation}
\label{eq:scaledham}
\frac{\hat{H}}{K}=-\eta \hat{n}+\frac{1}{N}\hat{n}(\hat{n}-1)-\chi\hat{P}_{2}
\end{equation}
where $N = N_{\rm Fock} - 1$
and $\chi = \xi / N$. At the critical point of a phase transition, all
properties are expected to scale as a power law $N^{-A}$, where $A$ is
the scaling exponent. Particularly important is the order parameter,
$\nu =\left\langle \hat{n}_{0}(\chi)\right\rangle /N$, which scales as
$\nu(\chi_{\rm c}) =A_{n0}N^{-A_{n1}}$. The order parameter as a function of $\chi$
is shown in figure \ref{fig:fig13} for $\eta =-1$ and different values of $N$. For the Hamiltonian operator
with truncated Hilbert space $\mathcal{H}$ of dimension $N_{\rm Fock}=N+1$, it is possible
to carry out calculations with large values of $N$ and thus identify the
critical point and scaling behavior accurately. The phase transition is 2nd order since,
at the critical point $\chi_{\rm c} = 0.5$, the first derivative of the order parameter $\nu$
is discontinuous (Ehrenfest classification).

The scaling behavior of the order parameter for the $\eta =-1$ case is $\nu (\chi_{\rm c})=0.2065\times N^{-0.6365}$. A similar situation occurs for $\eta = -2$ and $\eta = 0$. The critical value is $\chi_{\rm c} = -\frac{1}{2}\eta$. Note that the critical point $\chi_{\rm c}$ is not equal to the kissing point $\xi_{\rm k}$, rather, they are related by $\chi_{\rm c} = \frac{1}{4} \xi_{\rm k}$.

\subsection{QPT and ESQPT of the Liouville superoperator} \label{liouqpts}

The Liouville superoperator of (\ref{eq:squeezedkerrliou}) contains three control parameters,
$\eta =\omega /K,\xi =\varepsilon _{2}/K$ and $\zeta =\kappa /K$, the latter
being the ratio of the dissipator, $\kappa $, to the Kerr coefficient, $K$.
Minganti et al.~\cite{minganti} provided a formal definition of a
QPT of order $M$ for open systems, similar to the Ehrenfest classification of
QPTs of Hamiltonian operators mentioned in section~\ref{hamqpts},
\begin{equation}
\label{eq:qpt}
\lim_{\xi \rightarrow \xi _{\rm c}}\left\vert \frac{\partial ^{M}}{\partial \xi
^{M}}\lim_{N\rightarrow \infty }Tr\left[ \hat{\rho}_{ss}(\xi ,N)\hat{o}\right] \right\vert =+\infty .
\end{equation}
In this definition, the ground state of $\hat{H}$, $E_{0}$, is replaced by
the steady-state density matrix $\hat{\rho}_{ss}$,
which is a normalized eigenmatrix of the Liouvillian corresponding to a zero eigenvalue,
$\lambda_{0} = 0$, $\hat{\rho}_{ss} = \hat{\rho}_{0} / \mathrm{Tr}\left[ \hat{\rho}_{0} \right]$.
While in the case of closed systems one needs to consider the eigenvalues of $\hat{H}$ as in
figures \ref{fig:fig11} and \ref{fig:fig12}, for open systems one must consider
the eigenvalues $\lambda _{i}$ of the Liouville superoperator $\mathcal{L}$, ordered in such a way that
$\left\vert \mathrm{Re}\left[ \lambda _{0}\right] \right\vert <\left\vert \mathrm{Re}[\lambda _{1}]\right\vert <...<\left\vert \mathrm{Re}[\lambda_{n}]\right\vert $, and then look at their properties.

\subsubsection{Second order QPT of the squeezed Kerr oscillator.} \label{kerr2ndorderqpt}

The study of 2nd order QPT of the Liouville superoperator is
straightforward, as one needs to consider only the first non-zero eigenvalue 
$\lambda _{1}$. The real part $\mathrm{Re}[\lambda _{1}]$ (Liouvillian gap),
called also the asymptotic decay rate \cite{kessler}, is of importance since
it determines the slowest relaxation time of the system in the long-time limit,
$T_{X}$ (\ref{eq:relaxtime}).The imaginary part $\mathrm{Im}[\lambda_{1}]$,
called the Hamiltonian gap, is also relevant, since at the critical
point it closes \cite{kessler}, \cite{horstmann} and the levels touch.

\begin{figure}[ht!]
    \centering
    \includegraphics[width=0.7\textwidth]{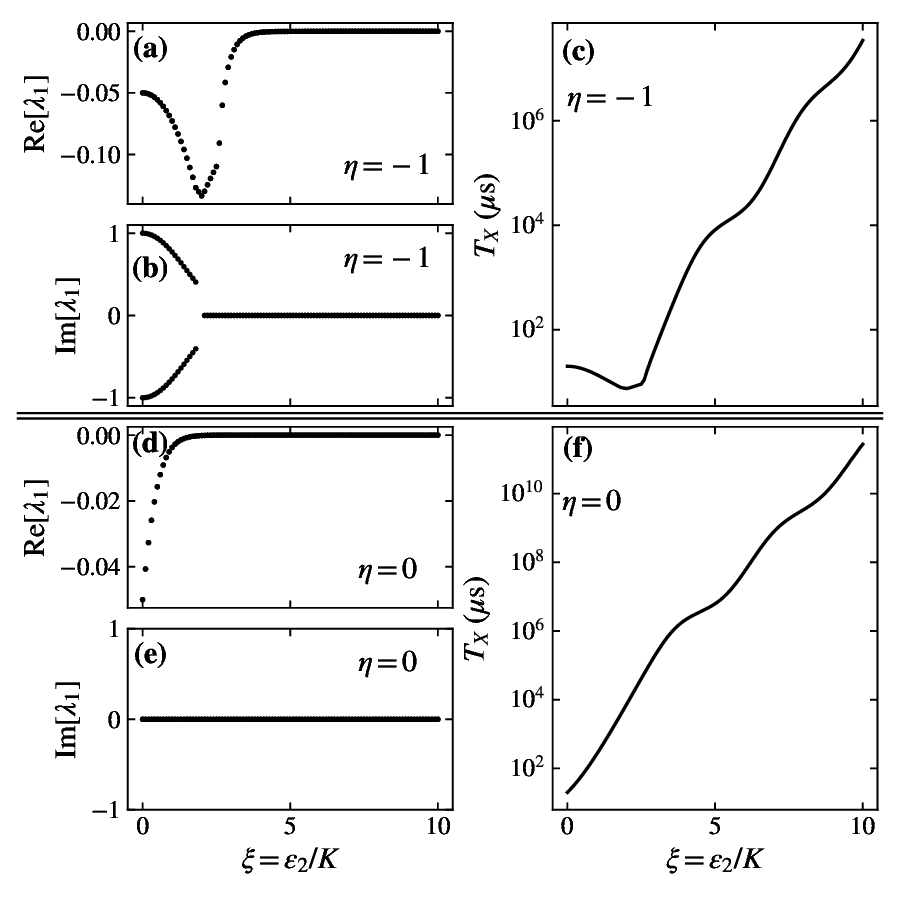}
    \caption{Top: (a) $\mathrm{Re}[\lambda _{1}]$ and (b) $\mathrm{Im}[\lambda _{1}]$
as a function of $\xi $ for the squeezed quadratic oscillator. (c) The
relaxation time $T_{X}$ as a function of $\xi $ for the squeezed quadratic
oscillator. Bottom: (d) $\mathrm{Re}[\lambda _{1}]$ and (e) $\mathrm{Im}[\lambda
_{1}]$ as a function of $\xi $ for the squeezed Kerr oscillator. (f) The
relaxation time $T_{X}$ as a function of $\xi $ for the squeezed Kerr
oscillator.}
    \label{fig:fig14}
\end{figure}

To illustrate these properties, we consider now two cases, $\eta =-1$ and $\eta =0$.
Consider first the case of $\eta =-1$, called the squeezed quadratic oscillator, with Hamiltonian
\begin{equation}
\label{eq:squeezedquadraticosc}
\frac{\hat{H}}{K}=\hat{n}^{2}-\xi \hat{P}_{2}
\end{equation}
and Liouville superoperator as in (\ref{eq:squeezedkerrliou}). A similar case was also
considered in \cite{minganti}, but with quadratic dissipation also included. The real and
imaginary parts of the eigenvalue $\lambda _{1}$ as a function of $\xi$ for $N=100$ are
shown in figure \ref{fig:fig14}a,b. One can see clearly the occurrence of a phase transition.
In $\mathrm{Im}[\lambda_{1}]$ the second order QPT appears as the closing of the
Hamiltonian gap at $\xi_{\rm k} = 2$. In $\mathrm{Re}[\lambda_{1}]$ it appears as a
minimum. The relaxation time $T_{X}$ is shown in figure \ref{fig:fig14}c. Here the QPT
appears as a minimum at $\xi _{\rm k}=2$.

Consider next the case $\eta =0$, called the squeezed Kerr oscillator,
with Hamiltonian
\begin{equation}
\label{eq:squeezedkerrosc2}
\frac{\hat{H}}{K}=\hat{n}(\hat{n}-1)-\xi \hat{P}_{2}.
\end{equation}
The real and imaginary parts of the eigenvalue $\lambda _{1}$ as a function of $\xi$
for $N=100$ are shown in figure \ref{fig:fig14}d,e. The QPT here is not seen because it is obscured by the fact that the
kissing point is at $\xi _{\rm k}=0$. The relaxation time is shown in figure \ref{fig:fig14}f.

\begin{figure}[ht!]
    \centering
    \includegraphics[width=0.7\textwidth]{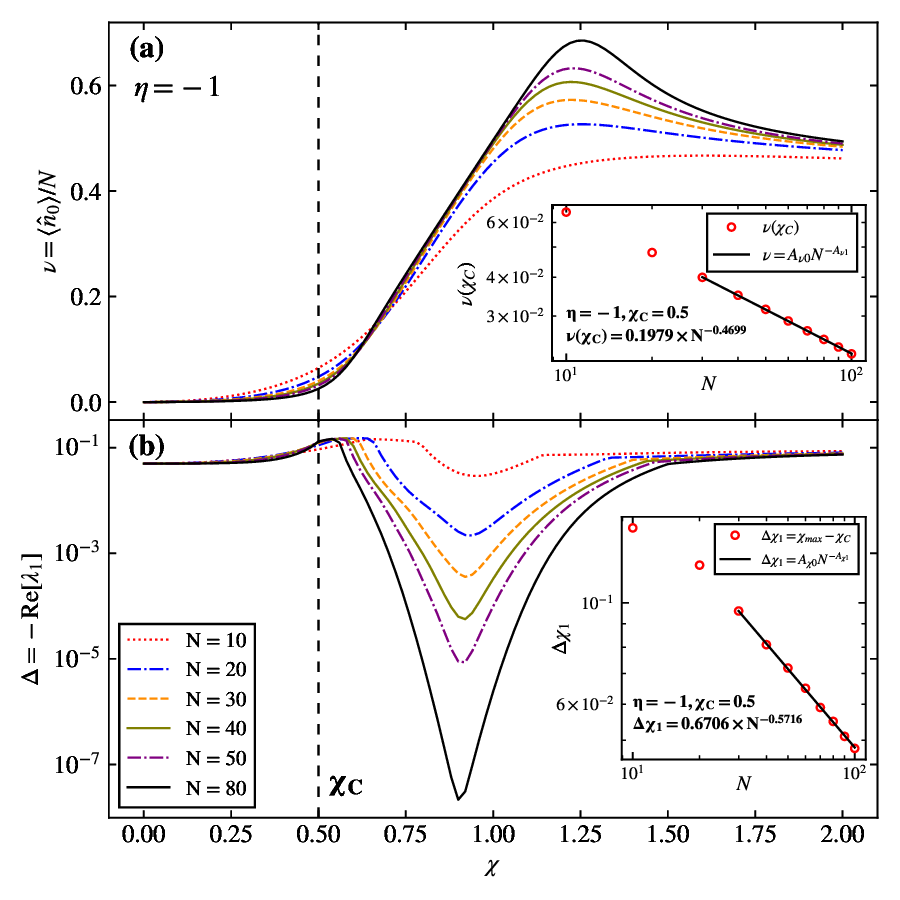}
    \caption{(a) The order parameter $\nu =\left\langle \hat{n}_{0}\right\rangle / N$
    as a function of $\chi$ for $\zeta =0.1$;
(b) The Liouvillian gap $\Delta =-\mathrm{Re}[\lambda _{1}]$ as function of $\chi$
for $\zeta =0.1$. Both in (a) and (b) $10\leq N\leq 80$. The critical value $\chi_{\rm c}$ is shown by a dashed vertical line. The scaling behavior of $\nu (\chi_{\rm c})$ is plotted in the insert in (a), and the scaling behavior of $\Delta \chi_{1}=\chi _{\max }-\chi_{\rm c}$ is plotted in the insert in (b).}
    \label{fig:fig15}
\end{figure}

We now consider the scaling properties of the squeezed quadratic oscillator, $\eta =-1$, with
scaled Hamiltonian as in (\ref{eq:scaledham}) and linear dissipator of (\ref{eq:squeezedkerrliou}).
Scaling properties now depend also on $\zeta =\kappa /K$ and one needs to consider both
the order parameter $\nu =\left\langle \hat{n}_{0}\right\rangle / N$ and the Liouvillian gap,
$\Delta = -\mathrm{Re}[\lambda _{1}]$. While the former is a property of the Hamiltonian
and appears in section~\ref{hamqpts}, the latter appears only for dissipative phase transitions.
These quantities are shown in figure \ref{fig:fig15}, as a function of $\chi$ for a fixed value of
$\zeta =0.1$. A similar calculation of the order parameter was done previously in \cite{minganti}.

\begin{figure}[ht!]
    \centering
    \includegraphics[width=0.7\textwidth]{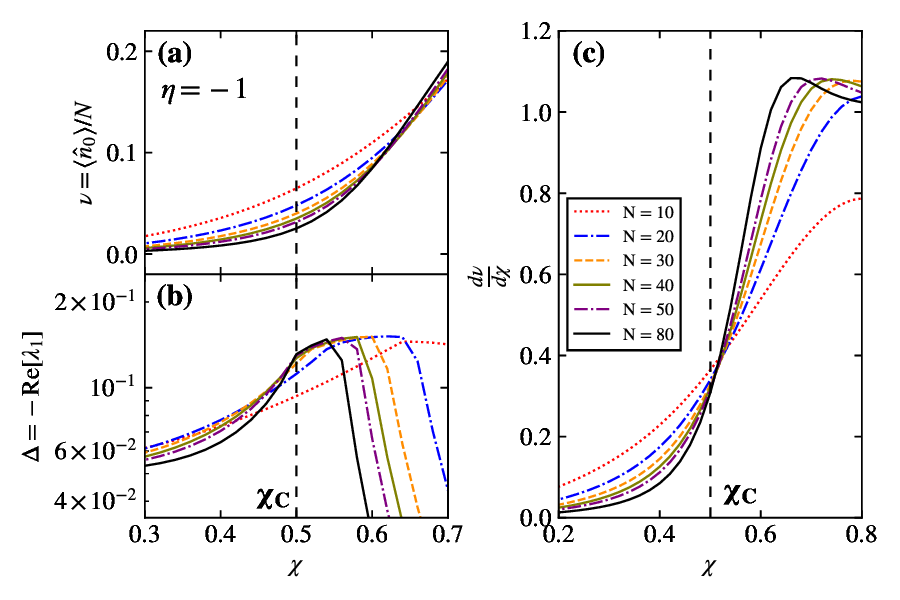}
    \caption{An enlarged version of the region around the critical point $\chi_{c}$
of figure \ref{fig:fig15}. (a) Order parameter; (b) Liouvillian gap; (c) First derivative of the order parameter,
$\frac{\rm{d} \nu}{\rm{d} \chi}$, plotted to emphasize its discontinuous behavior near $\chi_{c}$.}
    \label{fig:fig16}
\end{figure}

We first note that the Hilbert space $\mathcal{H}\otimes \mathcal{H}$ of the
Liouvillian is of dimension $N_{\rm Fock}\times N_{\rm Fock}$,
and therefore we consider values $N\leq 80$ for computational purposes,
in contrast with studies of the Hamiltonian.
We also note that in figure \ref{fig:fig15}a the critical value
of the order parameter $\nu $ is the same as for the
Hamiltonian for $\eta =-1$, $\chi_{\rm c}=0.5$, and that the transition is of 2nd order
since the slope of $\nu$ is discontinuous at $\chi_{\rm c}$
in the $N \rightarrow \infty$ limit.
This property was also emphasized in a recent preprint where experimental results were presented \cite{beaulieu}.
The scaling exponent, shown in the insert of
figure \ref{fig:fig15}a is however different from that of the Hamiltonian,
$\nu (\chi_{\rm c})=0.1979\times N^{-0.4699}$,
indicating a dependence on $\zeta $ (here $\zeta =0.1$).
As the discontinuity here is smoothed out by finite $N$ effects,
in figure \ref{fig:fig16} we display a close up view of $\nu$, $\Delta$, and $\frac{\rm{d} \nu}{\rm{d} \chi}$
around the critical point to better illustrate behavior at $\chi_{\rm c}$ and the discontinuity in $\frac{\rm{d} \nu}{\rm{d} \chi}$.
Another feature apparent in figure \ref{fig:fig15}a is the maximum occurring at about $\chi \cong 1.25$
before $\nu$ approaches its asymptotic value of $\nu =1/2$. This
maximum is also related to finite $N$ effects and is a result of
approximating a piecewise discontinuous function by a continuous one,
sometimes called the Gibbs phenomenon \cite{walker}.

Of particular interest is the behavior of the Liouvillian gap $\Delta =-\mathrm{Re}\left[ \lambda _{1}\right]$
in figure \ref{fig:fig15}b. At the critical point $\chi _{c}=0.5$, $\Delta$
approaches a maximum value in the $N \rightarrow \infty$ limit,
as expected from the previous figure \ref{fig:fig14}a for the unscaled system,
where $\mathrm{Re}\left[ \lambda _{1}\right]$ has a minimum.
This behavior is apparent in figure \ref{fig:fig16}b.
Several quantities can be used to describe the scaling behavior of $\Delta $.
We use here the quantity $\Delta \chi _{1}=\chi _{\max }-\chi _{c}$, where 
$\chi _{\max }$ is the value of $\chi $ at the maximum value attained by $
\Delta $, e.g. $\Delta \left( \chi _{\max }\right) =\max \left[ \Delta 
\right] $. The scaling behavior of this quantity is shown in the insert of
figure \ref{fig:fig15}b, as $\Delta \chi _{1}=0.6706\times N^{-0.5716}$.

Another important feature of figure \ref{fig:fig15}b is the minimum of
$\Delta$ occurring at $\chi_{\min }\cong 0.9$.
This minimum, found at large values of $\chi=\xi /N$,
was not noted in \cite{minganti}, since the calculation was stopped at values
less than $\chi _{\min }$. The minimum occurs at ultrastrong values of the
squeezing strength $\xi _{\min }\cong 0.9N$, which diverge as $N\rightarrow
\infty $. The study of this ultrastrong regime is outside the scope of the
present article.

\subsubsection{First order QPT of the squeezed Kerr oscillator.} \label{kerr1storderqpt}

The study of first order QPTs, both non-dissipative and dissipative, is more complicated than that of second order. For the Hamiltonian operator, first order QPTs were investigated in complex algebraic models \cite{carr} either, years ago, in models in large numbers of dimensions such as the $su(6)$ interacting boson model \cite{iac6}, or, more recently, in coupled Bose-Fermi systems such as the Dicke or Rabi Models \cite{shen,yang}.

\begin{figure}[ht!]
    \centering
    \includegraphics[width=0.7\textwidth]{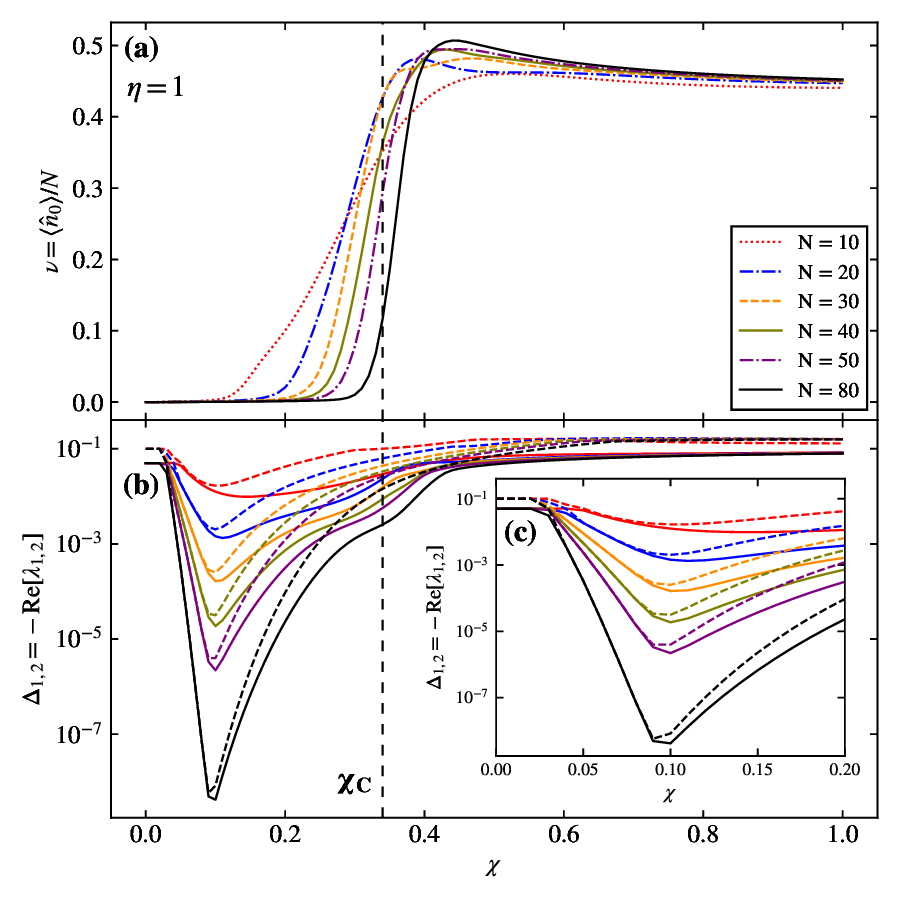}
    \caption{(a) The order parameter $\nu $ as a function of $\chi$ for $\zeta =0.1$;
(b) The Liouvillian gaps $\mathrm{Re}[\lambda _{1}]$ (full line), 
$\mathrm{Re}[\lambda _{2}]$(dashed line), as a function of $\chi $ for $\zeta=0.1$;
(c) close-up of the Liouvillian gap for small $\xi \leq 0.2$. The dashed vertical line
indicates the value of $\chi_{\rm c}$ at $N=80$.}
    \label{fig:fig17}
\end{figure}

In the case of higher-dimensional models \cite{iac1} it was found \cite{iac6} that one needs more than one excited eigenvalue, $E_{1}$, to study 1st order transitions. The complications found in the study of 1st order QPTs of Hamiltonian operators are even more apparent in the study of 1st order QPTs of Liouville superoperators. In \cite{minganti}, it was suggested to consider two non-zero eigenvalues $\lambda_{1}$ and $\lambda_{2}$ of the Liouville superoperator.
Here, we consider specifically the case of $\eta = +1$, with scaled Hamiltonian of (\ref{eq:scaledham}) and linear dissipator of (\ref{eq:squeezedkerrliou}), and study its order parameter $\nu $ and the real parts of its eigenvalues $\mathrm{Re}[\lambda_{1}]$, $\mathrm{Re}[\lambda _{2}]$. These quantities are shown in figure \ref{fig:fig17} as
a function of $\chi$ for $\zeta =0.1$.

In figure \ref{fig:fig17}a one can clearly see the discontinuity in the order parameter $\nu$
(a property of the lowest eigenstate) indicating a first order QPT (Ehrenfest classification).
As $N$ increases, the value $\chi_{\rm c}$ at
which the discontinuity occurs increases. At $N=80$ it is $\chi_{\rm c}\cong
0.34$. We estimate by extrapolation that in the asymptotic limit, $N\rightarrow \infty $, $\chi_{\rm c}=0.4$.
The behavior of the Liouvillian gap is more complicated.
Both $\Delta _{1} = -\mathrm{Re}[\lambda _{1}]$ and $\Delta _{2} = -\mathrm{Re}[\lambda _{2}]$ have a minimum.
Up to $\chi \cong 0.10$, the values of $\Delta _{1}$ and $\Delta _{2}$ are
degenerate. From that point on, they split. The first order QPT which occurs
for the order parameter $\nu$ at $\chi_{\rm c}\cong 0.34$ (for $N=80$), appears in the
Liouvillian gap as an inflection point in $\Delta _{1}$, observed in figure \ref{fig:fig17}b.
The properties of the Liouvillian gap for small values of $\chi$ observed in figure \ref{fig:fig17}b-c
are consistent with panel (e) of figure \ref{fig:fig8}. We also note that the behavior of the
Liouvillian gap as a function of $\chi$, while similar to that of \cite{minganti} for small
$\chi \le 0.1$, differs for $\chi \ge 0.1$. The difference may be due to the fact that the authors
in \cite{minganti} included also a quadratic dissipator, and that we take the asymptotic
thermodynamic limit to be $N\rightarrow \infty$ where $N$ is the maximum number
of bosons, $N = N_{\rm Fock} - 1$.

\section{Temperature dependence} \label{tempdependence}

In this section we consider the temperature dependence of the eigenvalues of
the Liouville superoperator $\mathcal{L}$ and the relaxation time $T_{X}$.

\subsection{Harmonic oscillator} \label{qhotempdep}

\begin{figure}[ht!]
    \centering
    \includegraphics[width=0.5\textwidth]{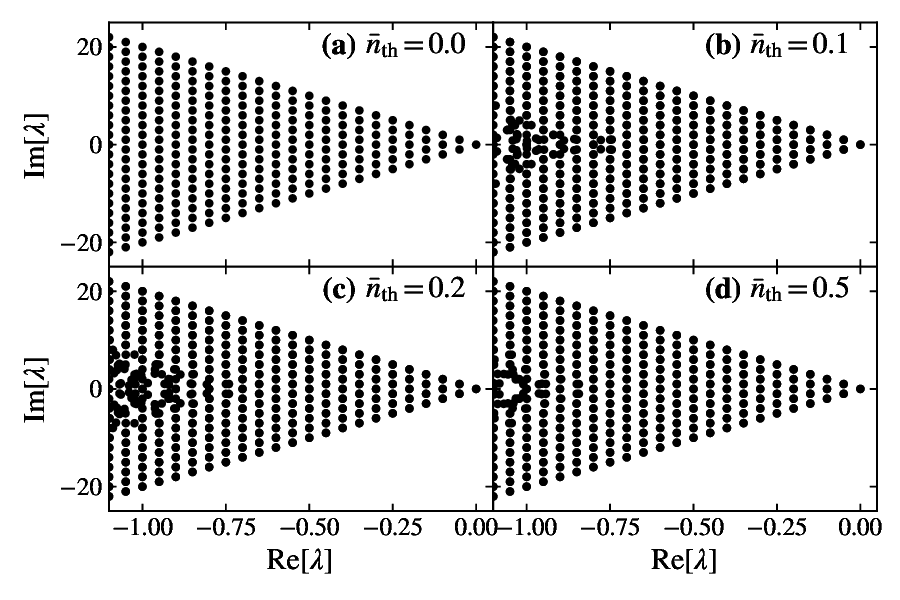}
    \caption{Temperature dependence of the eigenvalue spectrum of the harmonic
oscillator with $N_{\rm Fock}=120$ for $\bar{n}_{\rm th}=0.0,0.1,0.2,0.5$.}
    \label{fig:fig18}
\end{figure}

For the harmonic oscillator, $\omega =-1,$ $K=0$, the spectrum of eigenvalues of
the Liouville superoperator is shown in figure \ref{fig:fig18} for $N_{\rm Fock}=120$ and
average temperatures $\bar{n}_{\rm th}=0,0.1,0.2,0.5$. The spectrum of the harmonic
oscillator Liouvillian is independent of of temperature, a surprising result that can be derived
using the presence of a weak $u(1)$ symmetry and third quantization methods \cite{mcdonald}.
It is important to note that numerical calculations
converge to the analytic result for large $N_{\rm Fock}$ values, while for small $N_{\rm Fock}$ values,
truncation and finite $N$ effects play an important role.

\subsection{Kerr oscillator} \label{kotempdep}

\begin{figure}[ht!]
    \centering
    \includegraphics[width=0.5\textwidth]{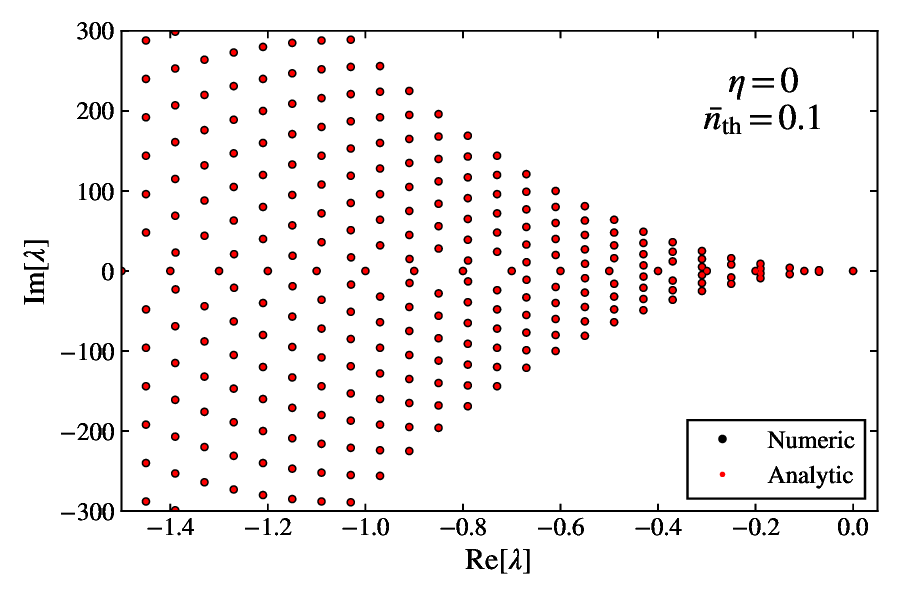}
    \caption{Spectrum of the Liouvillian of the Kerr oscillator with $\omega = 0,$ $K=1,$
    $\varepsilon _{2}=0,$ $\kappa = 0.1,$ for $\bar{n}_{\rm th}=0.1$ and $N_{\rm Fock}=80$.
    Red points are obtained analytically using the presence of a weak
    $u(1)$ symmetry and third quantization \cite{mcdonald}.}
    \label{fig:fig19}
\end{figure}

The spectrum of the Kerr Liouvillian for $\omega = 0,$ $K=1,$ $\varepsilon _{2}=0,$
$\kappa = 0.1,$ $\bar{n}_{\rm th}=0.1,$ $N_{\rm Fock}=80$ is shown in figure \ref{fig:fig19}.
It is a deformed version of the spectrum at zero temperature shown in figure \ref{fig:fig2}b.
This spectrum agrees with the analytic derivation of \cite{mcdonald}. Due to the Kerr nonlinearity,
$\left( K \neq 0 \right)$, numerical calculations converge to analytic results for smaller
$N_{\rm Fock}$ than for the harmonic oscillator.

\begin{figure}[ht!]
    \centering
    \includegraphics[width=0.9\textwidth]{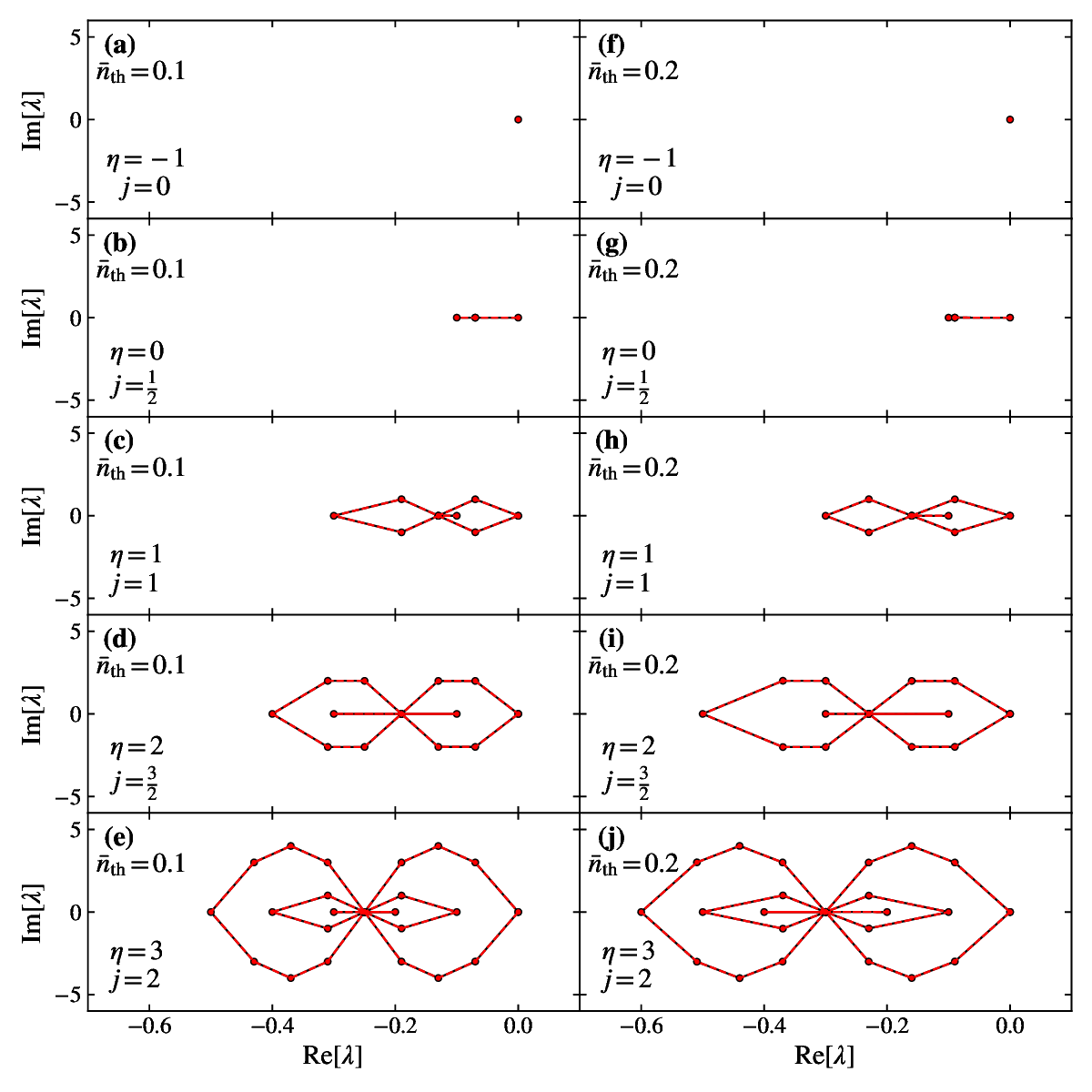}
    \caption{Temperature dependence of the eigenvalue spectrum of the
    Liouvillian of the Kerr oscillator with $N_{\rm Fock}=80$. Red points are obtained
    analytically via third quantization \cite{mcdonald}}
    \label{fig:fig20}
\end{figure}

It is of interest to study the effect of temperature on the eigenvalues of the Liouvillian for the
quasi-spin representation $\left\vert j,m_{j}\right\rangle$ of figure \ref{fig:fig4} when going from
$\bar{n}_{\rm th}=0$ to $\bar{n}_{\rm th}=0.2$. This is shown in figure \ref{fig:fig20}, for
$\omega =-1,0,1,2,3,$ $K=1,$ $\varepsilon _{2}=0$ and a small value of $N_{\rm Fock}=10$. It appears
that the geometric double ellipsoidal structure of the eigenvalues persists, although somewhat deformed and
with the degeneracy of the middle point lifted. The structure of the results of figure \ref{fig:fig20} is
confirmed by the analytic formula of \cite{mcdonald}.

\subsection{Squeezed Kerr oscillator} \label{squeezedkotempdep}

\begin{figure}[ht!]
    \centering
    \includegraphics[width=0.7\textwidth]{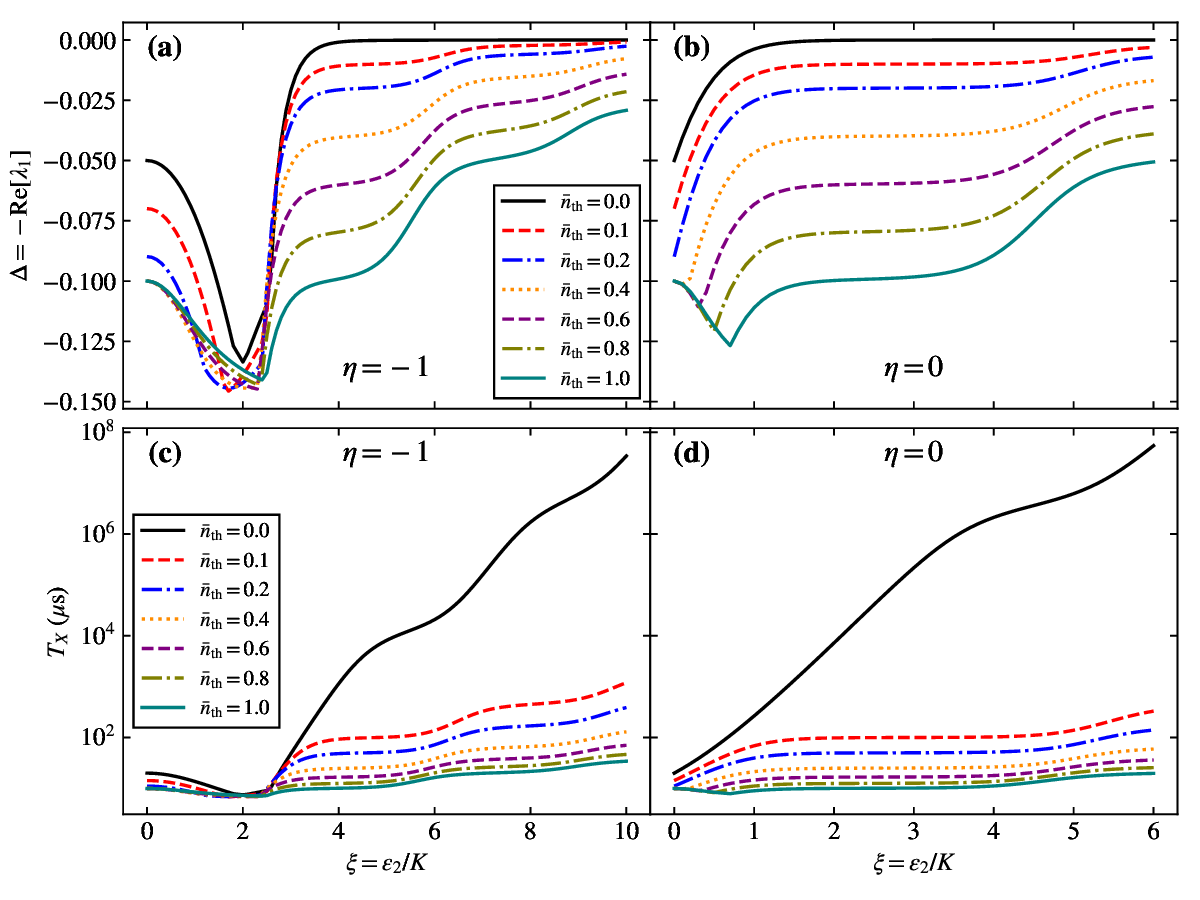}
    \caption{Behavior of $\mathrm{Re}[\lambda _{1}]$ as a function of $\xi $ for
several values of $\bar{n}_{\rm th}$ and $\eta =-1$ (a)$,\eta =0$
(b). Behavior of $T_{X}$ as a function of $\xi $ for several values of 
$\bar{n}_{\rm th}$ and $\eta =-1$ (c), $\eta =0$ (d).}
    \label{fig:fig21}
\end{figure}

\begin{figure}[ht!]
    \centering
    \includegraphics[width=0.7\textwidth]{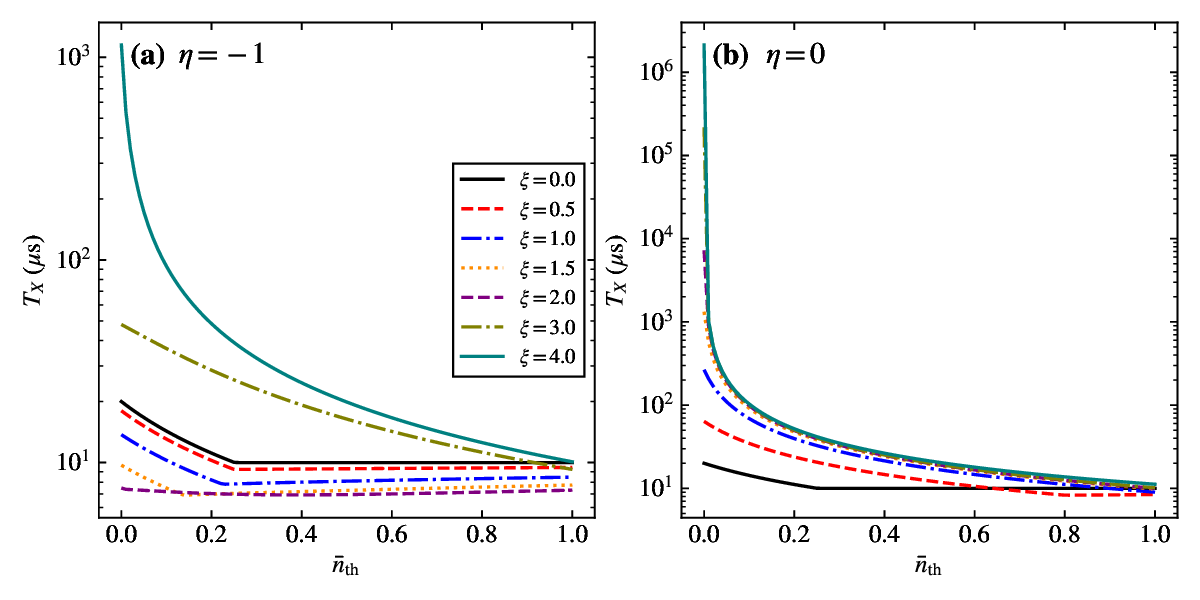}
    \caption{Behavior of $T_{X}$ as a function of $0\leq \bar{n}_{\rm th}\leq 1.0$
for several values of $\xi $ and $\eta =-1$ (a)$,\eta =0$ (b).}
    \label{fig:fig22}
\end{figure}

\begin{figure}[ht!]
    \centering
    \includegraphics[width=0.7\textwidth]{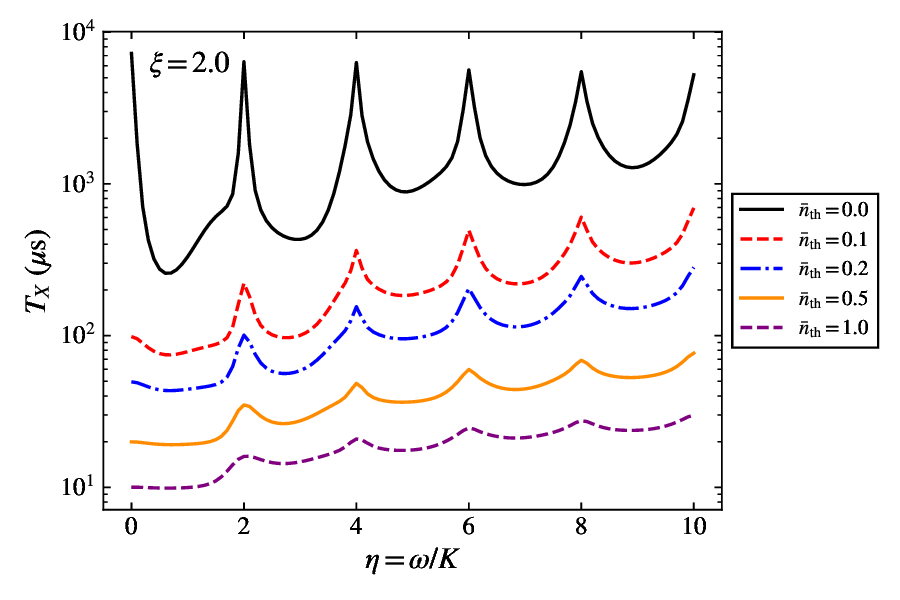}
    \caption{Behavior of $T_{X}$ as a function of $\eta =\omega /K$ for $\xi =2$
and several values of $\bar{n}_{\rm th}$. In this figure, $\kappa = 0.1$ $\mu$s$^{-1}$.}
    \label{fig:fig23}
\end{figure}

Here we present results for the squeezed Kerr oscillator, $\eta = \omega / K =-1,0,$ $\xi = \varepsilon _{2} / K \neq 0$,
with $N_{\rm Fock}=80$ to ensure convergence of lowest lying eigenvalues.
The behavior of the relaxation time $T_{X}$ as a function of $\bar{n}_{\rm th}$
for several values of $\xi $ and $\eta $ was investigated in great detail in 
\cite{venkatraman2}. Here we show in figure \ref{fig:fig21}a-b the behavior of $\mathrm{Re}[\lambda _{1}]$
as a function of $\xi $ for several values of  $\bar{n}_{\rm th}$
between $0.0$ and $1.0$ and $\eta =-1,0$. One can clearly see the second
order dissipative QPT occurring at $\xi_{\rm k} \cong 2.0$ for $\eta =-1$ and a
smoother behavior for $\eta =0$ where the QPT occurs at $\xi =0$ and it is
thus masked when $\xi >0$. From $\mathrm{Re}[\lambda _{1}]$, we can calculate
the relaxation time $T_{X}=-\frac{1}{\mathrm{Re}[\lambda _{1}]}$ as a function
of $\xi $, given in figure \ref{fig:fig21}c-d. Here, we see
a dramatic decrease of the relaxation time even for small $\bar{n}_{\rm th}=0.1$. This
behavior is emphasized in the subsequent figure \ref{fig:fig22}, which shows the relaxation
as a function of $0.0\leq \bar{n}_{\rm th}\leq 1.0$ for $\eta =-1$ and $\eta=0$.
A similar situation occurs for the relaxation time as a function of $\eta $ \cite{venkatraman2},
shown in figure \ref{fig:fig23} for $\xi_{\rm k} =2.0$ and several
values of $\bar{n}_{\rm th}$ between $0.0$ and $1.0$. One can see here that
although the peaked structure for $\eta = \mathrm{even}$ persists even at large
temperatures, it is gradually washed away. These results stress
that while the squeezed Kerr oscillator can be parametrically tuned to maximize
relaxation time, thermal effects and maintaining low temperatures are equally important
to doing so.

\section{Summary and conclusions} \label{conclusions}

In this article, we have investigated the symmetries of the Liouville
superoperator, $\mathcal{L}$, of one-dimensional parametric oscillators,
especially the squeeze-driven Kerr oscillator.
We have shown that for integer values of the ratio $\eta =\omega /K$
in the Kerr Hamiltonian $\hat{H}=-\omega \hat{n}+K\hat{n}(\hat{n}-1)$, the
spectrum of $\mathcal{L}$ has a characteristic double-ellipsoidal structure
and an hitherto unknown $su(2)$ quasi-spin symmetry, which reflects the symmetry of the
Hamiltonian $\hat{H}$. We have also shown that, as a result of this
quasi-spin symmetry, the relaxation time $T_{X}$ is
particularly large for even integer values of the ratio $\eta $ in the
squeeze-driven Hamiltonian $\hat{H}=-\omega \hat{n}+K\hat{n}(\hat{n}-1)-\varepsilon _{2}\hat{P}_{2}$,
a result of importance for the generation of long-lived states
useful in quantum computing. On the other hand, we have shown that
at nonzero temperature the relaxation time $T_{X}$ decreases dramatically,
even for low thermal populations $\bar{n}_{\rm th} \cong 0.1$.
Our combined results suggest that `optimal' Kerr devices are for
$\eta = \mathrm{even} = 4,6, ... , \xi = \rm{large} \geq 4.0$,
kept at the lowest possible temperature, $\bar{n}_{\rm th} \leq 0.1$.

The results presented here can be extended to oscillators with higher order
squeezing, cubic, $k_{\rm s}=3$, and quartic, $k_{\rm s}=4$, which can be realized
experimentally \cite{venkatraman1,venkatraman2} and to higher order
dissipation, such as quadratic, $k_{\rm d}=2$, and cubic, $k_{\rm d}=3$, which may play a
role in experiments \cite{venkatraman2}. Our results are also of
relevance to all parametric one-dimensional oscillators
and to all other models with Hamiltonian operators which can be cast in the
form of non-linear squeezed oscillators. This includes the Lipkin model, with
Hamiltonian
\begin{equation}
\label{eq:lipkinham}
\hat{H}=\omega \hat{J}_{z}+K\hat{J}_{z}^{2}-\varepsilon _{2}(\hat{J}_{+}^{2} + \hat{J}_{-}^{2}),
\end{equation}
and dissipators $\mathcal{D}[\hat{J}_{\pm
}]\hat{\rho}(t)$ \cite{rubio} which, together with the Kerr oscillator and
the one-dimensional vibron model, form a ``universality class'' of parametric
oscillators.

Our recognition that the Liouville superoperator of the Kerr oscillator has a
quasi-spin symmetry $su(2)$ reflecting the symmetry of its Hamiltonian
has major implications for the study of Open Parametric Oscillators
(OPO) \cite{goto1,goto2}. The study of symmetries of the Liouvillian
performed here can be extended to two coupled oscillators,
$su_{1}(2) \oplus su_{2}(2)$, in the same way in which is done in nuclear physics
for the proton-neutron interacting boson model,
$su_{1}(6) \oplus su_{2}(6)$ \cite{iac1}, and in molecular physics for triatomic molecules,
$su_{1}(4)\oplus su_{2}(4)$ \cite{iac2}, and most importantly, to a large number of
coupled oscillators on a lattice $\sum_{i}\oplus su_{i}(2)$, in the same way in which
it is done in the algebraic theory of crystal vibrations \cite{iac7,dietz}.

Finally, the study of symmetries and dissipative quantum phase transitions
of Liouvillian presented here can be extended to more complex models, such
as the Rabi, Dicke and Jaynes-Cummings models \cite{shen,yang}, the
Hamiltonian of which is expressed in terms of boson, $\hat{a}^{\dag },\hat{a}
$ and fermion, $\hat{\sigma}_{x},\hat{\sigma}_{y},\hat{\sigma}_{z}$,
operators and for which the quantum phase transitions of the Hamiltonian
have already been studied \cite{shen,yang}.

\section*{Acknowledgements} \label{acknowledgements}

F I acknowledges discussions with R G Corti\~nas on possible experimental detection of symmetries
of Liouvillians and F P\'erez-Bernal and L F Santos on QPT and ESQPT of the squeezed Kerr
oscillator. F I and C V C acknowledge L Viola for helping us clarify some aspects related to
the spectral properties of squeezed harmonic oscillator Liouvillians. C V C acknowledges
University Fellowship support from the
Yale University Physics Department.

\appendix

\section{Review of the quasi-spin symmetry $\mathbf{su(2)}$}  \label{appendixa}

\begin{figure}[ht!]
    \centering
    \includegraphics[width=0.7\textwidth]{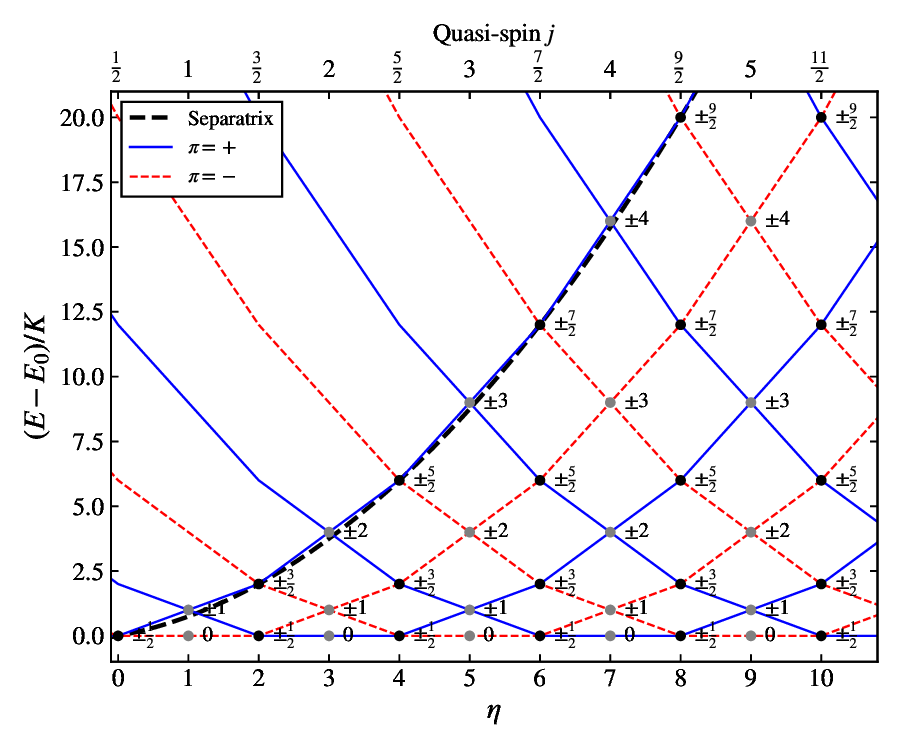}
    \caption{Excitation energy of the Hamiltonian $\hat{H}_{1}$ (\ref{eq:apxscaledham}) as a
function of the parameter $\eta =\omega /K$ showing the degeneracies that
occur for integer values of the parameter $\eta $ and the classification of
states in terms of quasi-spin quantum numbers $\left\vert j,m_{j}\right\rangle$. States are colored by parity, and the separatrix, marked with a black dashed line, $E_{s} / K = \eta / 2 + \eta^{2} / 4$, divides the spectrum into two phases.}
    \label{fig:figA1}
\end{figure}

Consider the dimensionless Hamiltonian
\begin{equation}
\label{eq:apxscaledham}
\hat{H}_{1}=\frac{\hat{H}}{K}=-\left( \eta +1\right) \hat{n}+\hat{n}
^{2}=-\eta ^{\prime }\hat{n}+\hat{n}^{2}
\end{equation}
The eigenvalues of $\hat{H}_{1}$ counted from the lowest state are shown in
figure \ref{fig:figA1}.

To the left of the dashed line, called the separatrix, states are singly
degenerate with eigenvalues $E=-\eta ^{\prime }n+n^{2}$. To the right of the
separatrix and for $\eta ^{\prime }=\eta +1=\mathrm{integer}$, degeneracies
occur. The degenerate points can be characterized by quasi-spin quantum
numbers $\left\vert j,m_{j}\right\rangle $. The values of the quasi-spin are
$j=\frac{\eta ^{\prime }}{2}=\frac{\eta +1}{2}$, while those of $m_{j}$ are
\begin{eqnarray}
\label{eq:apxquasispin}
m_{j} &=&\pm j,\pm \left( j-1\right) ,...,\pm 1/2;\quad j=\mathrm{half-integer}; \quad \eta=\mathrm{even}  \nonumber \\
m_{j} &=&\pm j,\pm \left( j-1\right) ,...,0; \quad j=\mathrm{integer}; \quad \eta=\mathrm{odd}.
\end{eqnarray}
The eigenvalues $E_{m}$ are given by
\begin{equation}
\label{eq:apxenergytotal}
E_{m} =m_{j}^{2}- j^{2} .
\end{equation}
When counted from the lowest state, they can be written as
\begin{eqnarray}
\label{eq:apxenergy}
E_{m} &=&m_{j}^{2}-1/4\quad j=\mathrm{half-integer}  \nonumber \\
E_{m} &=&m_{j}^{2} \qquad j=\mathrm{integer}.
\end{eqnarray}
Both sets of eigenvalues correspond to the quasi-spin symmetry $su(2)\supset so(2)$.
To further elucidate this quasi-spin symmetry, it is convenient to
explicitly construct the representations $\left\vert j,m_{j}\right\rangle $ with two boson
operators $\hat{b}_{1},\hat{b}_{2}$ with eigenvalues of the number operators 
$\hat{n}_{1},\hat{n}_{2}$ satisfying $n_{1}+n_{2}=N$.
In this situation, products of the two boson operators, $\hat{b}_{1}^{\dag} \hat{b}_{2}$,
$\hat{b}_{2}^{\dag} \hat{b}_{2}$, $\hat{n}_{1} = \hat{b}_{1}^{\dag} \hat{b}_{1}$,
$\hat{n}_{2} = \hat{b}_{2}^{\dag} \hat{b}_{2}$, form the Lie algebra $u(2)$, and the degenerate states of the Hamiltonian
$\hat{H}_{1}$ can be realized by letting $\hat{n} = \hat{n}_{1}$ in (\ref{eq:apxscaledham}) and $N = \eta ^{\prime} = 2j$.
The minimum energy is found to be $E_{\rm min} = -\eta ^{\prime 2} / 4 + 1/4$ for $\eta ^{\prime} = \mathrm{odd}$ and 
$E_{\rm min} = -\eta ^{\prime 2} / 4$ for $\eta ^{\prime} = \mathrm{even}$, yielding energies counted from the lowest state,
\begin{eqnarray}
\label{eq:apxenergyu2}
E_{n} - E_{\rm min} &=& n_{1}^{2} - \eta ^{\prime} n_{1} + \eta ^{\prime 2} / 4 - 1/4 \quad \eta ^{\prime} = \mathrm{odd} \nonumber \\
E_{n} - E_{\rm min} &=& n_{1}^{2} - \eta ^{\prime} n_{1} + \eta ^{\prime 2} / 4  \qquad \eta ^{\prime} = \mathrm{even}.
\end{eqnarray}
These formulae can be converted to the $su(2)$ quasi-spin notation, $\left\vert j,m_{j}\right\rangle $, by noting $n_{1}+n_{2}=N=\eta ^{\prime} = 2j$,
and using,
\begin{equation}
\label{eq:apxsu2labels}
j=\left( \frac{n_{2}+n_{1}}{2}\right) ,\quad m_{j}=\left( \frac{n_{2}-n_{1}}{2}\right)
\end{equation}
to explicitly obtain (\ref{eq:apxenergy}) from (\ref{eq:apxenergyu2}). Furthermore, we can compactly
write the Hamiltonian $\hat{H}_{1}$, counting from the lowest energy state, in the $N=2j$ subspace of the full Hilbert space in terms of quasi-spin operators,
\begin{equation}
\label{eq:apxquasispinham}
\hat{H}_{1} = \hat{j}_{z}^{2} - \frac{1}{8} \left( 1 - (-1)^{2j} \right)
\end{equation}
where $\hat{j}_{z} \left\vert j,m_{j}\right\rangle = m_{j} \left\vert j,m_{j}\right\rangle$. The quasi-spin
dynamic symmetry of $\hat{H}_{1}$ is evident here as it is written exclusively in terms of the
$su(2)\supset so(2)$ invariant Casimir operator $\hat{j}_{z}$.

In terms of the two boson operators $\hat{b}_{1},\hat{b}_{2}$, the wave
functions of the degenerate states can be written as
\begin{equation}
\label{eq:apxbosonstate}
\left\vert n_{1},n_{2}\right\rangle =\frac{1}{\sqrt{n_{1}!\left(N-n_{1}\right) !}}
\left( \hat{b}_{2}^{\dag }\right)^{N-n_{1}}\left( \hat{b}_{1}\right)^{n_{1}}
\left\vert 0\right\rangle
\end{equation}
The notation $\left\vert n_{1},n_{2}\right\rangle $ can be converted to the
usual quasi-spin notation by means of (\ref{eq:apxsu2labels}), giving
\begin{equation}
\label{eq:apxspinstate}
\left\vert j,m_{j}\right\rangle =\frac{1}{\sqrt{\left( j-m_{j}\right) !\left(j+m_{j}\right) !}}
\left( \hat{b}_{2}^{\dag }\right)^{j+m_{j}}\left( \hat{b}_{1}^{\dag }\right)^{j-m_{j}}
\left\vert 0\right\rangle .
\end{equation}
Additional details can be found in \cite{iac3} and \cite{iac5}.

\section*{References} \label{references}
\bibliography{bibliography}

\end{document}